\documentclass[twocolumn]{aastex701}

\usepackage{amsmath} 
\usepackage{color}
\definecolor{black}{rgb}{0,0,0}
\definecolor{blue}{rgb}{0.2,0.2,0.81}
\definecolor{orange}{rgb}{0.92,0.33,0.16}
\definecolor{gray}{rgb}{.5,0.5,0.5}

\begin{document}

\title{The 3D Cosmic Shoreline for Nurturing Planetary Atmospheres}

\shorttitle{Probabilistic Cosmic Shorelines}

\author[orcid=0000-0002-3321-4924]{Zach K. Berta-Thompson}
\affiliation{University of Colorado Boulder, Department of Astrophysical and Planetary Sciences}
\email[show]{zach.bertathompson@colorado.edu}

\author[orcid=0000-0001-6484-7559]{Patcharapol Wachiraphan}
\affiliation{University of Colorado Boulder, Department of Astrophysical and Planetary Sciences}
\email{patcharapol.wachiraphan@colorado.edu}

\author[orcid=0000-0001-8504-5862]{Catriona Murray}
\affiliation{University of Colorado Boulder, Department of Astrophysical and Planetary Sciences}
\email{catriona.murray@colorado.edu}

\newcommand{\mrb}{-0.025 \pm 0.03}
\newcommand{\mrbjustvalue}{-0.025}
\newcommand{\mrbsigma}{-1.22 \pm 0.087}
\newcommand{\mrbsigmajustvalue}{-1.22}
\newcommand{\mrm}{3.37 \pm 0.03}
\newcommand{\mrmjustvalue}{3.37}
\newcommand{\mrmsigma}{-0.224 \pm 0.035}
\newcommand{\mrmsigmajustvalue}{-0.224}
\newcommand{\mrscatteratearth}{29.5 \pm 2.6}
\newcommand{\mrscatteratearthjustvalue}{29.5}
\newcommand{\mrscatteratsmall}{138_{-25}^{+33}}
\newcommand{\mrscatteratsmalljustvalue}{138}
\newcommand{\mrpowerRvesc}{0.843 \pm 0.011}
\newcommand{\mrpowerRvescjustvalue}{0.843}
\newcommand{\mrpowerMvesc}{2.84 \pm 0.011}
\newcommand{\mrpowerMvescjustvalue}{2.84}
\newcommand{\mrpowerp}{3.16 \pm 0.011}
\newcommand{\mrpowerpjustvalue}{3.16}

\newcommand{\solarnomagmanofreezelnw}{-2.93_{-2.3}^{+3}}
\newcommand{\solarnomagmanofreezelnwjustvalue}{-2.93}
\newcommand{\solarnomagmanofreezelogfo}{0.526_{-0.2}^{+1.1}}
\newcommand{\solarnomagmanofreezelogfojustvalue}{0.526}
\newcommand{\solarnomagmanofreezep}{1.14_{-0.43}^{+1.9}}
\newcommand{\solarnomagmanofreezepjustvalue}{1.14}
\newcommand{\solarnomagmanofreezeq}{0.325_{-1.1}^{+0.45}}
\newcommand{\solarnomagmanofreezeqjustvalue}{0.325}
\newcommand{\solarnomagmanofreezew}{0.053_{-0.048}^{+1}}
\newcommand{\solarnomagmanofreezewjustvalue}{0.053}
\newcommand{\solarnomagmanofreezefo}{3.36_{-1.3}^{+43}}
\newcommand{\solarnomagmanofreezefojustvalue}{3.36}
\newcommand{\solarnomagmanofreezelogLnohz}{-0.757_{-1.6}^{+3.1}}
\newcommand{\solarnomagmanofreezelogLnohzjustvalue}{-0.757}
\newcommand{\solarnomagmanofreezeLnohz}{0.175_{-0.17}^{+2.2e+02}}
\newcommand{\solarnomagmanofreezeLnohzjustvalue}{0.175}
\newcommand{\solarnomagmanofreezewninetyfive}{0.314_{-0.28}^{+6.2}}
\newcommand{\solarnomagmanofreezewninetyfivejustvalue}{0.314}
\newcommand{\solarnomagmanofreezewninetyfiveasfluxfactor}{2.06_{-0.98}^{+2.9e+06}}
\newcommand{\solarnomagmanofreezewninetyfiveasfluxfactorjustvalue}{2.06}
\newcommand{\solarnomagmanofreezeqfromard}{0.31_{-0.12}^{+0.67}}
\newcommand{\solarnomagmanofreezeqfromardjustvalue}{0.31}

\newcommand{\solaranylnw}{-1.37_{-0.62}^{+0.68}}
\newcommand{\solaranylnwjustvalue}{-1.37}
\newcommand{\solaranylogfo}{0.865_{-0.41}^{+0.56}}
\newcommand{\solaranylogfojustvalue}{0.865}
\newcommand{\solaranyp}{3.76_{-0.51}^{+0.73}}
\newcommand{\solaranypjustvalue}{3.76}
\newcommand{\solaranyq}{0 \pm 0.93}
\newcommand{\solaranyqjustvalue}{0}
\newcommand{\solaranyw}{0.254_{-0.12}^{+0.25}}
\newcommand{\solaranywjustvalue}{0.254}
\newcommand{\solaranyfo}{7.32_{-4.5}^{+19}}
\newcommand{\solaranyfojustvalue}{7.32}
\newcommand{\solaranylogLnohz}{-0.003 \pm 2.7}
\newcommand{\solaranylogLnohzjustvalue}{-0.003}
\newcommand{\solaranyLnohz}{0.993_{-0.99}^{+5.5e+02}}
\newcommand{\solaranyLnohzjustvalue}{0.993}
\newcommand{\solaranywninetyfive}{1.5_{-0.7}^{+1.5}}
\newcommand{\solaranywninetyfivejustvalue}{1.5}
\newcommand{\solaranywninetyfiveasfluxfactor}{31.4_{-25}^{+8.8e+02}}
\newcommand{\solaranywninetyfiveasfluxfactorjustvalue}{31.4}
\newcommand{\solaranyqfromard}{0.509_{-0.24}^{+0.33}}
\newcommand{\solaranyqfromardjustvalue}{0.509}

\newcommand{\allnomagmalnw}{-1.40_{-0.30}^{+0.32}}
\newcommand{\allnomagmalnwjustvalue}{-1.4}
\newcommand{\allnomagmalogfo}{2.62_{-0.31}^{+0.45}}
\newcommand{\allnomagmalogfojustvalue}{2.62}
\newcommand{\allnomagmap}{5.9_{-0.43}^{+0.61}}
\newcommand{\allnomagmapjustvalue}{5.9}
\newcommand{\allnomagmaq}{1.17_{-0.20}^{+0.28}}
\newcommand{\allnomagmaqjustvalue}{1.17}
\newcommand{\allnomagmaw}{0.247_{-0.065}^{+0.092}}
\newcommand{\allnomagmawjustvalue}{0.247}
\newcommand{\allnomagmafo}{414_{-2.1e+02}^{+7.6e+02}}
\newcommand{\allnomagmafojustvalue}{414}
\newcommand{\allnomagmalogLnohz}{-2.23_{-0.21}^{+0.18}}
\newcommand{\allnomagmalogLnohzjustvalue}{-2.23}
\newcommand{\allnomagmaLnohz}{0.006_{-0.002}^{+0.003}}
\newcommand{\allnomagmaLnohzjustvalue}{0.006}
\newcommand{\allnomagmawninetyfive}{1.46_{-0.38}^{+0.54}}
\newcommand{\allnomagmawninetyfivejustvalue}{1.46}
\newcommand{\allnomagmawninetyfiveasfluxfactor}{28.7_{-17}^{+71}}
\newcommand{\allnomagmawninetyfiveasfluxfactorjustvalue}{28.7}
\newcommand{\allnomagmaqfromard}{1.54_{-0.18}^{+0.27}}
\newcommand{\allnomagmaqfromardjustvalue}{1.54}

\newcommand{\allnomagmanofreezelnw}{-2.41_{-0.76}^{+0.67}}
\newcommand{\allnomagmanofreezelnwjustvalue}{-2.41}
\newcommand{\allnomagmanofreezelogfo}{2.73_{-0.35}^{+0.51}}
\newcommand{\allnomagmanofreezelogfojustvalue}{2.73}
\newcommand{\allnomagmanofreezep}{6.3_{-1.3}^{+1.7}}
\newcommand{\allnomagmanofreezepjustvalue}{6.3}
\newcommand{\allnomagmanofreezeq}{1.27_{-0.25}^{+0.37}}
\newcommand{\allnomagmanofreezeqjustvalue}{1.27}
\newcommand{\allnomagmanofreezew}{0.09_{-0.048}^{+0.085}}
\newcommand{\allnomagmanofreezewjustvalue}{0.09}
\newcommand{\allnomagmanofreezefo}{543_{-3e+02}^{+1.2e+03}}
\newcommand{\allnomagmanofreezefojustvalue}{543}
\newcommand{\allnomagmanofreezelogLnohz}{-2.14_{-0.26}^{+0.23}}
\newcommand{\allnomagmanofreezelogLnohzjustvalue}{-2.14}
\newcommand{\allnomagmanofreezeLnohz}{0.007_{-0.003}^{+0.005}}
\newcommand{\allnomagmanofreezeLnohzjustvalue}{0.007}
\newcommand{\allnomagmanofreezewninetyfive}{0.53_{-0.28}^{+0.5}}
\newcommand{\allnomagmanofreezewninetyfivejustvalue}{0.53}
\newcommand{\allnomagmanofreezewninetyfiveasfluxfactor}{3.38_{-1.6}^{+7.3}}
\newcommand{\allnomagmanofreezewninetyfiveasfluxfactorjustvalue}{3.38}
\newcommand{\allnomagmanofreezeqfromard}{1.61_{-0.21}^{+0.3}}
\newcommand{\allnomagmanofreezeqfromardjustvalue}{1.61}

\newcommand{\exonomagmalnw}{-2.8_{-1.8}^{+1}}
\newcommand{\exonomagmalnwjustvalue}{-2.8}
\newcommand{\exonomagmalogfo}{2.72_{-0.44}^{+0.65}}
\newcommand{\exonomagmalogfojustvalue}{2.72}
\newcommand{\exonomagmap}{6.37_{-2.2}^{+3.1}}
\newcommand{\exonomagmapjustvalue}{6.37}
\newcommand{\exonomagmaq}{1.29_{-0.28}^{+0.41}}
\newcommand{\exonomagmaqjustvalue}{1.29}
\newcommand{\exonomagmaw}{0.061_{-0.051}^{+0.11}}
\newcommand{\exonomagmawjustvalue}{0.061}
\newcommand{\exonomagmafo}{522_{-3.3e+02}^{+1.8e+03}}
\newcommand{\exonomagmafojustvalue}{522}
\newcommand{\exonomagmalogLnohz}{-2.14 \pm 0.38}
\newcommand{\exonomagmalogLnohzjustvalue}{-2.14}
\newcommand{\exonomagmaLnohz}{0.007_{-0.004}^{+0.01}}
\newcommand{\exonomagmaLnohzjustvalue}{0.007}
\newcommand{\exonomagmawninetyfive}{0.358_{-0.3}^{+0.63}}
\newcommand{\exonomagmawninetyfivejustvalue}{0.358}
\newcommand{\exonomagmawninetyfiveasfluxfactor}{2.28_{-1.1}^{+7.4}}
\newcommand{\exonomagmawninetyfiveasfluxfactorjustvalue}{2.28}
\newcommand{\exonomagmaqfromard}{1.6_{-0.26}^{+0.38}}
\newcommand{\exonomagmaqfromardjustvalue}{1.6}

\newcommand{\solarnomagmalnw}{-1.37_{-0.63}^{+0.69}}
\newcommand{\solarnomagmalnwjustvalue}{-1.37}
\newcommand{\solarnomagmalogfo}{0.867_{-0.41}^{+0.56}}
\newcommand{\solarnomagmalogfojustvalue}{0.867}
\newcommand{\solarnomagmap}{3.76_{-0.51}^{+0.73}}
\newcommand{\solarnomagmapjustvalue}{3.76}
\newcommand{\solarnomagmaq}{-0.002 \pm 0.94}
\newcommand{\solarnomagmaqjustvalue}{-0.002}
\newcommand{\solarnomagmaw}{0.254_{-0.12}^{+0.25}}
\newcommand{\solarnomagmawjustvalue}{0.254}
\newcommand{\solarnomagmafo}{7.37_{-4.5}^{+19}}
\newcommand{\solarnomagmafojustvalue}{7.37}
\newcommand{\solarnomagmalogLnohz}{0.012 \pm 2.8}
\newcommand{\solarnomagmalogLnohzjustvalue}{0.012}
\newcommand{\solarnomagmaLnohz}{1.03_{-1}^{+6.1e+02}}
\newcommand{\solarnomagmaLnohzjustvalue}{1.03}
\newcommand{\solarnomagmawninetyfive}{1.5_{-0.7}^{+1.5}}
\newcommand{\solarnomagmawninetyfivejustvalue}{1.5}
\newcommand{\solarnomagmawninetyfiveasfluxfactor}{31.3_{-25}^{+9.3e+02}}
\newcommand{\solarnomagmawninetyfiveasfluxfactorjustvalue}{31.3}
\newcommand{\solarnomagmaqfromard}{0.51_{-0.24}^{+0.33}}
\newcommand{\solarnomagmaqfromardjustvalue}{0.51}

\newcommand{\allanylnw}{-1.05_{-0.26}^{+0.28}}
\newcommand{\allanylnwjustvalue}{-1.05}
\newcommand{\allanylogfo}{3.25_{-0.41}^{+0.55}}
\newcommand{\allanylogfojustvalue}{3.25}
\newcommand{\allanyp}{6.67_{-0.58}^{+0.77}}
\newcommand{\allanypjustvalue}{6.67}
\newcommand{\allanyq}{1.54_{-0.27}^{+0.35}}
\newcommand{\allanyqjustvalue}{1.54}
\newcommand{\allanyw}{0.351_{-0.08}^{+0.11}}
\newcommand{\allanywjustvalue}{0.351}
\newcommand{\allanyfo}{1.77e+03_{-1.1e+03}^{+4.5e+03}}
\newcommand{\allanyfojustvalue}{1.77e+03}
\newcommand{\allanylogLnohz}{-2.11_{-0.2}^{+0.16}}
\newcommand{\allanylogLnohzjustvalue}{-2.11}
\newcommand{\allanyLnohz}{0.008_{-0.003}^{+0.004}}
\newcommand{\allanyLnohzjustvalue}{0.008}
\newcommand{\allanywninetyfive}{2.07_{-0.47}^{+0.67}}
\newcommand{\allanywninetyfivejustvalue}{2.07}
\newcommand{\allanywninetyfiveasfluxfactor}{117_{-78}^{+4.3e+02}}
\newcommand{\allanywninetyfiveasfluxfactorjustvalue}{117}
\newcommand{\allanyqfromard}{1.91_{-0.24}^{+0.32}}
\newcommand{\allanyqfromardjustvalue}{1.91}

\newcommand{\exoanylnw}{-1.95_{-2}^{+0.67}}
\newcommand{\exoanylnwjustvalue}{-1.95}
\newcommand{\exoanylogfo}{3.73_{-0.47}^{+0.59}}
\newcommand{\exoanylogfojustvalue}{3.73}
\newcommand{\exoanyp}{7.68_{-3.9}^{+4.4}}
\newcommand{\exoanypjustvalue}{7.68}
\newcommand{\exoanyq}{1.86_{-0.3}^{+0.43}}
\newcommand{\exoanyqjustvalue}{1.86}
\newcommand{\exoanyw}{0.143_{-0.12}^{+0.14}}
\newcommand{\exoanywjustvalue}{0.143}
\newcommand{\exoanyfo}{5.36e+03_{-3.5e+03}^{+1.6e+04}}
\newcommand{\exoanyfojustvalue}{5.36e+03}
\newcommand{\exoanylogLnohz}{-2_{-0.41}^{+0.36}}
\newcommand{\exoanylogLnohzjustvalue}{-2}
\newcommand{\exoanyLnohz}{0.01_{-0.006}^{+0.013}}
\newcommand{\exoanyLnohzjustvalue}{0.01}
\newcommand{\exoanywninetyfive}{0.839_{-0.73}^{+0.8}}
\newcommand{\exoanywninetyfivejustvalue}{0.839}
\newcommand{\exoanywninetyfiveasfluxfactor}{6.91_{-5.6}^{+36}}
\newcommand{\exoanywninetyfiveasfluxfactorjustvalue}{6.91}
\newcommand{\exoanyqfromard}{2.19_{-0.28}^{+0.35}}
\newcommand{\exoanyqfromardjustvalue}{2.19}

\newcommand{\exonomagmanofreezelnw}{-2.65_{-1.8}^{+0.92}}
\newcommand{\exonomagmanofreezelnwjustvalue}{-2.65}
\newcommand{\exonomagmanofreezelogfo}{2.85_{-0.45}^{+0.66}}
\newcommand{\exonomagmanofreezelogfojustvalue}{2.85}
\newcommand{\exonomagmanofreezep}{6.04_{-2.1}^{+2.9}}
\newcommand{\exonomagmanofreezepjustvalue}{6.04}
\newcommand{\exonomagmanofreezeq}{1.31_{-0.28}^{+0.42}}
\newcommand{\exonomagmanofreezeqjustvalue}{1.31}
\newcommand{\exonomagmanofreezew}{0.071_{-0.059}^{+0.11}}
\newcommand{\exonomagmanofreezewjustvalue}{0.071}
\newcommand{\exonomagmanofreezefo}{700_{-4.5e+02}^{+2.5e+03}}
\newcommand{\exonomagmanofreezefojustvalue}{700}
\newcommand{\exonomagmanofreezelogLnohz}{-2.19 \pm 0.37}
\newcommand{\exonomagmanofreezelogLnohzjustvalue}{-2.19}
\newcommand{\exonomagmanofreezeLnohz}{0.006_{-0.004}^{+0.009}}
\newcommand{\exonomagmanofreezeLnohzjustvalue}{0.006}
\newcommand{\exonomagmanofreezewninetyfive}{0.415_{-0.35}^{+0.62}}
\newcommand{\exonomagmanofreezewninetyfivejustvalue}{0.415}
\newcommand{\exonomagmanofreezewninetyfiveasfluxfactor}{2.6_{-1.4}^{+8.3}}
\newcommand{\exonomagmanofreezewninetyfiveasfluxfactorjustvalue}{2.6}
\newcommand{\exonomagmanofreezeqfromard}{1.67_{-0.27}^{+0.39}}
\newcommand{\exonomagmanofreezeqfromardjustvalue}{1.67}

\newcommand{\applylnw}{-1.4_{-0.3}^{+0.32}}
\newcommand{\applylnwjustvalue}{-1.4}
\newcommand{\applylogfo}{2.62_{-0.31}^{+0.45}}
\newcommand{\applylogfojustvalue}{2.62}
\newcommand{\applyp}{5.9_{-0.43}^{+0.61}}
\newcommand{\applypjustvalue}{5.9}
\newcommand{\applyq}{1.17_{-0.2}^{+0.28}}
\newcommand{\applyqjustvalue}{1.17}
\newcommand{\applyw}{0.247_{-0.065}^{+0.092}}
\newcommand{\applywjustvalue}{0.247}
\newcommand{\applymercuryescapevelocity}{0.497_{-0.053}^{+0.047}}
\newcommand{\applymercuryescapevelocityjustvalue}{0.497}
\newcommand{\applymercuryrratio}{1.25_{-0.11}^{+0.10}}
\newcommand{\applymercuryrratiojustvalue}{1.25}
\newcommand{\applymercuryr}{0.478_{-0.042}^{+0.037}}
\newcommand{\applymercuryrjustvalue}{0.478}
\newcommand{\applyvenusflux}{264_{-1.3e+02}^{+4.6e+02}}
\newcommand{\applyvenusfluxjustvalue}{264}
\newcommand{\applyvenuslogflux}{2.42_{-0.3}^{+0.44}}
\newcommand{\applyvenuslogfluxjustvalue}{2.42}
\newcommand{\applyvenussemimajor}{0.062_{-0.024}^{+0.026}}
\newcommand{\applyvenussemimajorjustvalue}{0.062}
\newcommand{\applyearthlogluminositylimit}{-2.23_{-0.21}^{+0.18}}
\newcommand{\applyearthlogluminositylimitjustvalue}{-2.23}
\newcommand{\applyearthluminositylimit}{0.006_{-0.002}^{+0.003}}
\newcommand{\applyearthluminositylimitjustvalue}{0.006}
\newcommand{\applyearthandahalfluminositylimit}{0.001_{-0}^{+0.001}}
\newcommand{\applyearthandahalfluminositylimitjustvalue}{0.001}
\newcommand{\applyearthandahalflogluminositylimit}{-3.28_{-0.35}^{+0.3}}
\newcommand{\applyearthandahalflogluminositylimitjustvalue}{-3.28}
\newcommand{\applywninefivef}{1.46_{-0.38}^{+0.54}}
\newcommand{\applywninefivefjustvalue}{1.46}
\newcommand{\applywninefivev}{0.246_{-0.057}^{+0.073}}
\newcommand{\applywninefivevjustvalue}{0.246}
\newcommand{\applywninefiveL}{1.24_{-0.26}^{+0.33}}
\newcommand{\applywninefiveLjustvalue}{1.24}
\newcommand{\applyprobabilityLTTonefourfourfiveAc}{11.1_{-6.9}^{+12}}
\newcommand{\applyprobabilityLTTonefourfourfiveAcjustvalue}{11.1}
\newcommand{\applyprobabilityGJthreeninetwonineb}{21.8_{-12}^{+16}}
\newcommand{\applyprobabilityGJthreeninetwoninebjustvalue}{21.8}
\newcommand{\applyprobabilityLTTonefourfourfiveAb}{79.4_{-15}^{+11}}
\newcommand{\applyprobabilityLTTonefourfourfiveAbjustvalue}{79.4}
\newcommand{\applyprobabilityLHSoneonefourob}{99.9_{-0.81}^{+0.12}}
\newcommand{\applyprobabilityLHSoneonefourobjustvalue}{99.9}

\begin{abstract}

Various ``cosmic shorelines" have been proposed to delineate which planets have atmospheres. The fates of individual planet atmospheres may be set by a complex sea of growth and loss processes, driven by unmeasurable environmental factors or unknown historical events. Yet, defining population-level boundaries helps illuminate which processes matter and identify high-priority targets for future atmospheric searches. Here, we provide a statistical framework for inferring the position, shape, and fuzziness of an instellation-based cosmic shoreline, defined in the three-dimensional space of planet escape velocity, planet bolometric flux received, and host star luminosity. \added{We circumvent the need to estimate individual host stars' historical X-ray and extreme ultraviolet fluences by including luminosity in the definition of the shoreline, explicitly modeling how sharply such drivers of atmospheric escape intensify toward lower-luminosity M dwarf stars and marginalizing over the associated uncertainties.} Using Solar System and exoplanet atmospheric constraints, under the assumption that one planar boundary applies across a wide parameter space, we find the critical flux threshold for atmospheres scales with escape velocity with a power-law index of $p=\allnomagmap$, steeper than the canonical literature slope of $p=4$, and scales with stellar luminosity with a power-law index of $q=\allnomagmaq$, steep enough to disfavor atmospheres on Earth-sized planets out to the habitable zone for stars less luminous than $\log_{10} (L_\star/L_\sun) = \allnomagmalogLnohz$ (roughly spectral type M4V). \added{This model provides quantitative predictions for the probability any planet may have an atmosphere, which can be rigorously tested by upcoming JWST Rocky Worlds observations.} 

\end{abstract}


\section{Introduction} 
\label{s:introduction}

Where can atmospheres thrive? This question has grown more urgent as astronomers branch out from the Solar System to exoplanets, where atmospheres require great observational expense to measure or sometimes can only be imagined. A complete, precise, and predictive answer to this question might not exist, as each individual atmosphere is the integrated balance of difficult-to-model sources and sinks. Atmospheres grow through early accretion from primordial nebulae, through later impact delivery, through continual magmatic outgassing from the interior, and through evaporation or sublimation of surface volatiles. Atmospheres wither through myriad upper-atmosphere escape processes driven by stellar radiation, stellar winds, and/or impacts; through sequestering into the interior; and through condensation or deposition to the surface. These processes continuously interact with each other, they operate on timescales spanning minutes to gigayears, and they depend on historical environmental inputs that can be wildly uncertain, chaotic, or stochastic. On Earth and other inhabited planets, atmospheric evolution is further complicated by biogeochemical cycles that may include the influence of technological civilizations. For more on atmospheric evolution, see reviews by \citet{johnsonExospheresAtmosphericEscape2008, lammerAtmosphericEscapeEvolution2008, tianAtmosphericEscapeSolar2015b, owenAtmosphericEscapeEvolution2019a,  gronoffAtmosphericEscapeProcesses2020, wordsworthAtmospheresRockyExoplanets2022} and textbooks by \citet{chamberlainTheoryPlanetaryAtmospheres1987, pierrehumbertPrinciplesPlanetaryClimate2010, seagerExoplanetAtmospheresPhysical2010, ingersollPlanetaryClimates2013, lissauerFundamentalPlanetaryScience2019}.
 
Despite the incredible specifics needed to model an atmosphere's detailed history, we can still seek systematic trends among basic planet properties that may allow for the cultivation of an atmosphere. \citet[][hereafter ZC17]{zahnleCosmicShorelineEvidence2017a} distilled this idea into the search for a ``cosmic shoreline'', with dry volatile-poor atmosphereless worlds (the sand) on one side of the shoreline and worlds rich in volatiles or atmospheres on the other (the lake/sea/ocean). ZC17 explored log-linear boundaries in 2D spaces defined by planetary escape velocity $v_{\sf esc}$ -- a tracer of how strongly planets hold onto volatiles (or various combinations of $v_{\sf esc}$ with planet mass $M$, radius $R$, density $\rho$) -- and by various sources of incoming energy available to drive escape: the current bolometric flux\footnote{In this work we primarily use ``flux'' ($f$) to refer to the power per unit area (W/m$^2$) a planet receives from its star. It is equivalent to ``insolation'' (\underline{in}coming \underline{sol}ar radi\underline{ation}) as used by ZC17, ``instellation'' (\underline{in}coming \underline{stell}ar radi\underline{ation}) introduced for exoplanets by \citet{shieldsEffectHostStar2013}, or ``irradiance.''}
planets receive $f$, the cumulative X-ray and extreme ultraviolet (XUV) fluence planets have received over their history $F_{\sf XUV} = \int_{0}^{\sf now}f_{\sf XUV}(t) dt$, and/or the estimated velocity of giant impacts $v_{\sf imp}$. Although it is typically less than 0.01\% of a star's bolometric luminosity \citep{franceMUSCLESTreasurySurvey2016}, the difficult-to-measure XUV flux is distinctly important because it drives the upper-atmosphere heating and ionization that mediate many escape processes \citep{linskyInferringIntrinsicStellar2024}. \added{ Stellar XUV increases dramatically on the main-sequence toward lower mass/radius/luminosity stars \citep{2019_linsky_HostStarsTheir, wilsonMegaMUSCLESTreasurySurvey2025b, passRecedingCosmicShoreline2025}. We must account for this rising XUV when comparing planets around different host stars, especially since the planets that are easiest to observe are often transiting low-mass M dwarf stars \citep{blakeNearInfraredMonitoringUltracool2008, nutzmanDesignConsiderationsGroundBased2008a, morleyObservingAtmospheresKnown2017b}.} ZC17 identified $f \propto v_{\sf esc}^4$ and $F_{\sf XUV} \propto v_{\sf esc}^4$ as effective definitions of instellation-based cosmic shorelines, as well as $v_{\sf imp}/v_{\sf esc} = 5$ as a potential impact-driven shoreline \citep[see also][]{zahnleOriginsAtmospheres1998}.

The ZC17 instellation-based shorelines have been adopted among the exoplanet community trying to identify rocky exoplanets most likely to have atmospheres and to contextualize non-detections of such atmospheres from JWST \citep[][and references therein]{parkcoyPopulationlevelHypothesisTesting2024, 2025_ih_RockyPlanetsStars}. The Rocky Worlds STScI Director's Discretionary Time program is using 500 hours of JWST time to survey terrestrial transiting exoplanets for atmospheres \citep{redfieldReportWorkingGroup2024} and includes estimated location relative to the $F_{\sf XUV}$ shoreline as a metric for target prioritization\footnote{\href{https://rockyworlds.stsci.edu/}{rockyworlds.stsci.edu}}. Since the shoreline is being used, we want to help make it as useful as possible. 

In this work, we revisit the ZC17's instellation-based shorelines through the lens of Bayesian probabilistic modeling and incorporate new rocky exoplanet atmospheres contraints from JWST. We define a generative model for the probability of a planet having an atmosphere and use it to infer the location, slope, and width of a cosmic shoreline, along with uncertainties on these quantities. We expand the shoreline into 3D, using planetary escape velocity $v_{\sf esc}$, planetary bolometric flux $f$, and stellar luminosity $L_\star$ as three predictors for whether planets have atmospheres. The inclusion of stellar luminosity is designed to remove the need for star-by-star estimates of hard-to-measure environmental drivers for atmospheric escape (like high-energy fluence $F_{\sf XUV}$), moving them to where they can be modeled and marginalized more easily on an ensemble level. Thus, the predictors for atmospheres can stay rooted in easy-to-observe measurements, while still capturing trends in changing stellar environment toward lower mass stars \added{(such as increased $F_{\sf XUV}$ making them more capable of eroding planetary atmospheres)}. Acknowledging that a true underlying cosmic shoreline is likely crinkled with fjords and peninsulas, tidepools and islands, we apply this approximate model to explore the threshold for atmospheres on both a global scale (from hot transiting exoplanets to the outer edges of the Solar System as in ZC17) and local scale (only planets where CO$_2$ is likely to be in the gas phase) relevant to JWST's current detection capabilities and to habitability.

We assemble planet populations to analyze in \S\ref{s:data}, present the probability model and fitting methodology in \S\ref{s:fitting},  show the inferred shorelines in \S\ref{s:shorelines}, interpret the physical implications of the derived slopes in \S\ref{s:physics}, and conclude in \S\ref{s:conclusions}. Code to reproduce all plots in the paper and calculations are linked throughout with the \href{https://github.com/zkbt/shoreline}{\texttt{</>}} symbol.

\section{Curating the Data}
\label{s:data}

We assemble planetary properties using \texttt{exoatlas} \citep{berta-thompsonZkbtExoatlas2025}, a tool for accessing, filtering, and visualizing archival planet properties. Solar System data come from JPL Solar System Dynamics tables of major planets, dwarf planets, minor planets, and moons\footnote{\href{https://ssd.jpl.nasa.gov}{ssd.jpl.nasa.gov}}. Exoplanet data come from the NASA Exoplanet Archive's \citep{christiansenNASAExoplanetArchive2025}\footnote{\href{https://exoplanetarchive.ipac.caltech.edu}{exoplanetarchive.ipac.caltech.edu}} Planetary Systems Composite Parameters table \citep{nasaexoplanetscienceinstitutePlanetarySystemsComposite2020} which provides as many properties as possible for each planet, but sometimes combines values from independent and possibly inconsistent literature sources. Where necessary, \texttt{exoatlas} can pick specific references for particular properties from the larger Planetary System table \citep{nasaexoplanetscienceinstitutePlanetarySystemsTable2020}, which includes every published value for every planet. \added{Data were last retrieved on 27 February 2026 and spot-checked for outrageous planet parameter choices in the composite table for many of the key planets discussed below.}
In \texttt{exoatlas} all quantities have units attached with \texttt{astropy.units}, as well as uncertainties propagated through calculations with numerical samples using \texttt{astropy.uncertainty}. 

\subsection{What quantities do we use to predict atmospheres?}

For stellar luminosity $L_\star$, if not present in the raw table, \texttt{exoatlas} calculates it from stellar effective temperature $T_{\sf eff, \star}$ and stellar radius $R_\star$. For the average bolometric flux a planet receives $f = L_\star/(4\pi a^2)$, we first attempt to pull planet semimajor axis $a$ from the table, then, if $a$ is not present, we attempt to calculate it from the orbital period $P$ and stellar mass $M_\star$ via $P^2 = 4\pi^2 a^3/GM_\star$, and then finally, if necessary, from a transit-derived scaled semimajor axis ratio $a/R_\star$. For the gravitational escape velocity of the planet $v_{\sf esc}$, we calculate it as $v_{\sf esc} = \sqrt{2GM / R}$. However, many planets have measured radii but not masses, with either radial velocity wobbles or transit-timing variations too weak to detect (transiting planets) or no moons to provide dynamical masses (small Solar System objects). 

\begin{figure*}[ht!]
\includegraphics[width=\textwidth]{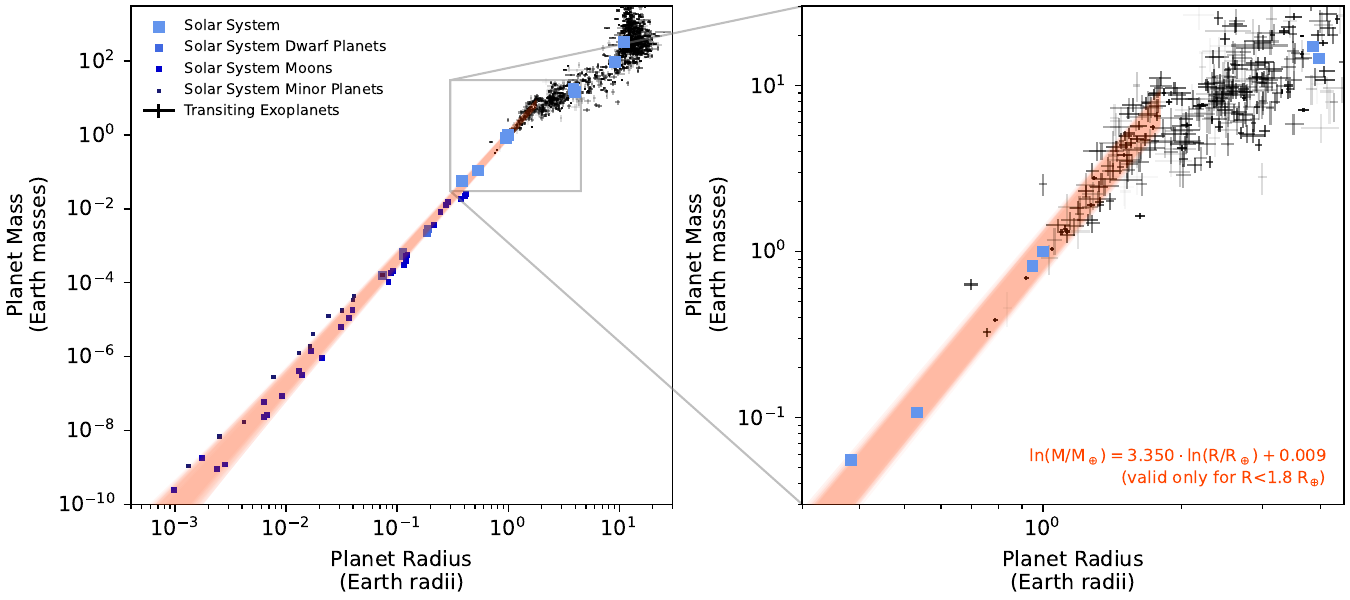}
\caption{To determine escape velocities for objects without measured masses, we derive an empirical mass-radius relationship from exoplanets (errorbars) and Solar System objects (squares). We use this relation, valid for rocky planets up to $1.8\rm{R}_\earth$, to estimate planet masses and uncertainties that incorporate the intrinsic scatter on the relation, the uncertainties on the model parameters, and uncertainties on the planet radii (\href{https://github.com/zkbt/shoreline/blob/main/notebooks/1-fit-mass-radius-relation.ipynb}{\texttt{</>}}).
}
\label{f:mass-radius}
\end{figure*}

To be able to include objects without measured masses in our analysis, we derive an empirical radius-to-mass relation from rocky objects with measured masses and radii. We limit to radii smaller than $1.8 {\rm R_\earth}$, as these are likely to be mostly terrestrial \citep{fultonCaliforniaKeplerSurveyVII2018, zengNewPerspectivesExoplanet2021, rogersMostSuperEarthsHave2025}. We fit a linear model $y = m\cdot x + b$ where we define $x_{\sf i} = \ln (R_{\sf i}/R_\earth)$ and $y_{\sf i} = \ln (M_{\sf i}/M_\earth)$, corresponding to a power-law relationship $M = CR^m$ where $C = e^b$. In addition to the measurement uncertainties on the data $\sigma_{x,i} = \sigma_{\ln R_{\sf i}} = \sigma_{R_{\sf i}}/{R_{\sf i}}$ and $ \sigma_{y_{\sf i}} =  \sigma_{\ln M_{\sf i}} =  \sigma_{M,i}/M_{\sf i}$, we include an intrinsic scatter on the relation $\sigma_{y}$. We allow this intrinsic scatter to vary with radius as $\ln \sigma_{y} = m_\sigma \cdot x + b_\sigma$ to capture the diversity of densities that grows toward very small objects due to effects of composition, structure, and porosity. We infer the parameters of this model ($m$, $b$, $m_\sigma$, $b_\sigma$) following a blog post by \citet[][see also \citealt{hoggDataAnalysisRecipes2010a}]{foreman-mackeyFittingPlaneData2017} with a Gaussian likelihood that analytically marginalizes over the uncertainties in both $x$ and $y$ and an uninformative prior on the slopes $P(m) \propto (1+m^2)^{-3/2}$ as in \citet{vanderplasFrequentismBayesianismPythondriven2014a}. We sample the posterior using \texttt{numpyro} \citep{phanComposableEffectsFlexible2019} with the No U-Turns Sampler \citep[NUTS;][]{hoffmanNoUTurnSamplerAdaptively2011}, using 4 chains each with 5,000 warm-up steps and 50,000 samples, reaching an \citet{gelmanInferenceIterativeSimulation1992} statistic of $\hat{R}=1.0$ and a bulk effective sample size $>10,000$ \citep[][see also \citealt{hoggDataAnalysisRecipes2018}]{vehtariRankNormalizationFoldingLocalization2021} for all parameters. Figure \ref{f:mass-radius} shows the result. The inferred slope of $m = \mrm $ is slightly steeper than a constant density ($m=3$) as expected due to self-gravity more strongly compressing larger planets, and the intercept $b = \mrb$ is close to Earth-like ($b=0$). The slope $m$ is similar to but slightly lower than other mass-radius relations for rocky planets:  3.58 \citep[$=1/0.279$;][]{chenPROBABILISTICFORECASTINGMASSES2017}, 3.45 \citep{otegiRevisitedMassradiusRelations2020}, 3.70 \citep[$=1/0.27$;][]{mullerMassradiusRelationExoplanets2024}. For the intrinsic scatter, the slope $m_\sigma = \mrmsigma $ and intercept $b_\sigma = \mrbsigma$ imply a \mrscatteratearthjustvalue\% scatter at 1 ${\rm R_\earth}$ that grows to \mrscatteratsmalljustvalue\% scatter at $10^{-3}$ ${\rm R_\earth}$. We incorporate the sample means and covariance matrix (which describe the nearly multivariate normal posterior well) into \texttt{exoatlas} to calculate mass estimates with uncertainties that include the uncertainties on the parameters themselves, the intrinsic scatter, and the input radius uncertainties. This relation is valid only for planets without gaseous envelopes contributing significantly to their overall size.

\subsection{What planets do we label as having atmospheres?}

To include in our probabilistic fit we label planets (\href{https://github.com/zkbt/shoreline/blob/main/notebooks/2-curate-and-label-planets.ipynb}{\texttt{</>}}) as having an atmosphere ($A_{\sf i} = 1$) or not ($A_{\sf i} = 0$). Planets with inconclusive or unmeasured atmospheres remain unlabeled ($A_{\sf i} = ?$) and are excluded from the fit. In both cases, to simplify the analysis, we exclude planets with quoted ages younger than 750 Myr, to minimize having to imagine the future states of still rapidly evolving planets \citep{lopezUnderstandingMassRadiusRelation2014a, chenEvolutionaryAnalysisGaseous2016b,  thaoFeatherweightGiantUnraveling2024b}.

We are generous in what we call ``having an atmosphere" ($A_{\sf i} = 1$), as in ZC17. For Solar System bodies, we include all major planets (everything except Mercury) and moons (Titan) with atmospheric surface pressures $>10^{-6}$ bar. We include outer Solar System moons or dwarf planets that have managed to retain significant volatile reservoirs \citep{schallerVolatileLossRetention2007}, either as seasonally sublimating atmospheres and/or substantial global N$_2$ and CH$_4$ volatile deposits on their surfaces: Triton, Pluto, Makemake, Eris \citep{youngStructureCompositionPlutos2018, sicardyConstraintsEvolutionTriton2024, grundyModerateRatiosMethane2024}. We label all other Solar System objects as $A_{\sf i} = 0$.

For exoplanets, planets larger than $2.2 \rm {R_\earth}$ are extremely difficult to explain with pure rocky compositions \citep{rogersMost16Earthradius2015b, zengNewPerspectivesExoplanet2021, rogersMostSuperEarthsHave2025}, so we label all planets 
with radii more than 1$\sigma$ over this limit as definitely requiring atmospheres (or significant volatiles) to explain their low densities $A_{\sf i} = 1$. Many planets smaller than this limit have atmospheres too, but we apply labels only to those with direct atmosphere measurements, as follows.

We apply $A_{\sf i} = 1$ to \added{two ultrahot small planets with evidence for atmospheres}. 55 Cnc e shows variable JWST eclipse spectra suggesting a (potentially stochastically outgassed) CO/CO$_2$ atmosphere \citep{huSecondaryAtmosphereRocky2024, patelSecondaryAtmosphereRocky2024}. \added{TOI-561 b shows cool dayside thermal emission, suggesting a thick atmosphere to redistribute heat \citep{2025_teske_ThickVolatileAtmosphere}; though this cool dayside may be alternatively explained by a very reflective Bond albedo \citep{mansfieldIdentifyingAtmospheresRocky2019b}, the planet's low density further strongly suggests a gaseous envelope \citep{2023_brinkman_TOI561LowdensityUltrashortperiod}.}

We apply $A_{\sf i} = 0$ to these rocky planets with eclipse observations of hot daysides that strongly suggest low albedos and poor global heat recirculation inconsistent with thick atmospheres \citep[see][]{kollIdentifyingCandidateAtmospheres2019c, mansfieldIdentifyingAtmospheresRocky2019b}. LHS 3844b, Gl 367b, TOI-1685b have a deep eclipses, symmetric phase curves, and dark night sides \citep{kreidbergAbsenceThickAtmosphere2019a, zhangGJ367bDark2024, 2025_luque_DarkBareRock}. GJ 1252b, TOI-1468b, LHS 1140c, \added{LTT 3780b}, and \added{GJ 3929b} have deep photometric eclipses \citep{crossfieldGJ1252bHot2022, meiervaldesHotRocksSurvey2025, fortuneHotRocksSurvey2025a, 2025_allen_HotRocksSurvey, 2025_xue_JWSTRockyWorlds}, and Gl 486b, GJ 1132b, and LTT 1445Ab have deep spectroscopic eclipses \citep{weinermansfieldNoThickAtmosphere2024, xueJWSTThermalEmission2024a, wachiraphanThermalEmissionSpectrum2025a}. \added{TRAPPIST-1b and TRAPPIST-1c also show deep thermal eclipses \citep{greeneThermalEmissionEarthsized2023, ihConstrainingThicknessTRAPPIST12023, ziebaNoThickCarbon2023}; although their atmosphere limits remain slightly ambiguous due to a tentative option of an inverted atmosphere on planet b \citep{ducrotCombinedAnalysis1282025} and systemic stellar uncertainties for both planets \citep{howardCharacterizingNearinfraredSpectra2023, rackhamRobustCorrectionsStellar2024,  fauchezStellarModelsAlso2025, limAtmosphericReconnaissanceTRAPPIST12023, radicaPromisePerilStellar2025, rathckeStellarContaminationCorrection2025}, we nonetheless label them both as effectively atmosphereless.} We caution that our labeling these exoplanets as $A_{\sf i} = 0$ does {\em not} mean planets necessarily have no atmosphere at all; for tidally locked planets, a JWST measurement of hot dayside emission might only constrain the atmospheric pressure on an individual planet to less than about 1-10 bar \citep{kollScalingAtmosphericHeat2022}, really saying simply that we have not detected a very thick Venus-like atmosphere.

We leave the following notable planets unlabeled ($A_{\sf i} = ?$), meaning they are ignored from the probabilistic fit, even if there have been suggestions about whether they have atmospheres. 
Kepler-10b and Kepler-78b have symmetric phase curves from Kepler, but with only the optical bandpass their deep eclipses are degenerate between reflected and thermal light, thus complicating atmospheric inferences \citep{sanchis-ojedaTransitsOccultationsEarthsized2013, estevesChangingPhasesAlien2015a, huSemianalyticalModelVisiblewavelength2015, singhProbingKeplersHottest2022}. K2-141 shows a K2 + Spitzer phase curve suggesting a high albedo or hot inversion layer, but whether an atmosphere is absolutely required remains uncertain \citep{singhProbingKeplersHottest2022, ziebaK2SpitzerPhase2022}. LHS 1478b and TOI-431b appear to have a shallow thermal eclipses, but more data are needed to rule out systematics \citep{augustHotRocksSurvey2025, monaghanLow45Mm2025a}. L 98-59 b's transmission spectrum may show tentative evidence of SO$_2$ \citep[possibly from tidally-heated volcanism;][] {seligmanPotentialMeltingExtrasolar2024} but is also consistent with featureless \citep{bello-arufeEvidenceVolcanicAtmosphere2025b}, so we leave it unlabeled. Otherwise, transmission spectra have not yet conclusively identified nor ruled out any rocky planet atmospheres due to degeneracies with clouds \citep{lustig-yaegerMirageCosmicShoreline2019} and/or stellar contamination \citep{mayDoubleTroubleTwo2023, moranHighTideRiptide2023}; we leave all transmission spectroscopy-based non-detections as $A_{\sf i} = ?$.

\added{\subsection{What do we expect atmospheres near the shoreline to be?}
Though ``having an atmosphere" can be partially agnostic to what makes up the atmosphere, it is worth commenting on the expected composition of planets near the shoreline. For close-in rocky planets where JWST is starting to provide atmospheric constraints, we focus on CO$_2$ (and to some extent O$_2$) as the likely major constituent of temperate atmospheres distilled by near-complete atmospheric erosion \citep{hamanoEmergenceTwoTypes2013a, lugerExtremeWaterLoss2015, schaeferPredictionsAtmosphericComposition2016c,  krissansen-tottonErosionLargePrimary2024}. Venus (92 bar) and Mars ($0.004-0.009$ bar) are both about 95\% CO$_2$ \citep{loddersPlanetaryScientistsCompanion1998}, and Earth too would likely have about 100 bars of atmospheric CO$_2$ (swamping our 1 bar of N$_2$ + O$_2$) if not for the presence of H$_2$O continually raining down, dissolving atmospheric CO$_2$, and locking it away into solid carbonates \citep{walkerNegativeFeedbackMechanism1981, lecuyerComparisonCarbonNitrogen2000, ingersollPlanetaryClimates2013, wordsworthAtmospheresRockyExoplanets2022, hansenDetectingAtmosphericCO22025}. Earth preserves its H$_2$O (and thus keeps our CO$_2$ trapped in limestone deposits) only because water precipitates before it reaches the upper atmosphere, where it would be dissociated by UV radiation and its H atoms lost to space. On warm Venus, the runaway greenhouse process \citep{komabayasi1ChengFen2XiangXinoDaQiQuantoShuiQuanwoYousuruJiaXiangDenaHuoXinggaYiDingnoTaiYangGuangXiadetoriuruBuLianSoknaPingHengWenDunituite1ChengFen2XiangXinoDaQiQuantoShuiQuanwoYousuruJiaXiangDenaHuoXinggaYiDingnoTaiYangGuangXiadetoriuruBuLianSoknaPingHengWenDunituiteDiscreteEquilibriumTemperatures1967, ingersollRunawayGreenhouseHistory1969} vaporizes water and drives it inevitably to the upper atmosphere, where its H escapes \citep{hamanoEmergenceTwoTypes2013a, leconteIncreasedInsolationThreshold2013, wordsworthWaterLossTerrestrial2013}. On cold Mars, hydrogen has also been preferentially lost \citep{jakoskyLossMartianAtmosphere2018}, with modern escape largely driven by dust dynamics carrying frozen water up to high altitudes \citep{chaffinMartianWaterLoss2021}. CO$_2$ (possibly mixed with O$_2$) atmospheres are also particular detectable for exoplanets with JWST, especially with Rocky Worlds' use of MIRI filter photometry centered on the strong 15 $\mu$m CO$_2$ absorption band \citep{morleyObservingAtmospheresKnown2017b, ihConstrainingThicknessTRAPPIST12023}.}

\added{
\subsection{What limits do we consider on the cosmic shoreline?}
\label{s:flux-limits}

We consider two thresholds in bolometric flux $f$ that we suspect could potentially matter when trying to define a continuous cosmic shoreline. First, we consider a {\em magma ocean} boundary that we set at $f_{\sf magma} = 1400 f_\oplus$, corresponding to a zero-albedo global equilibrium temperature of $T_{\sf magma} = 1700$K. Above roughly this instellation, significant surface melting may leave planets in a continuous magma ocean state \citep{boukareDeepTwophaseHemispherical2022}. At ultrahot temperatures, the vaporization of rock itself can generate significant atmospheres \citep{2009_schaefer_ChemistrySilicateAtmospheres, 2012_schaefer_VaporizationEarthApplication}. Molten planets can also provide direct gas exchange with a much deeper mantle volatile reservoir than cooler solidified planets, potentially replenishing long-living atmospheres even in the face of strong atmospheric erosion \citep{dornHiddenWaterMagma2021, huSecondaryAtmosphereRocky2024, 2025_teske_ThickVolatileAtmosphere}. Therefore, we will exclude planets hotter than $T_{\sf magma}$ from our main shoreline estimate.

Second, we consider a {\em CO$_2$ freezeout} boundary at $f_{\sf freeze} = 0.24f_\oplus$, corresponding to a temperature of $T_{\sf freeze} = 194$K (where the saturation vapor pressure of CO$_2$ drops below 1 bar; \citealt{pierrehumbertPrinciplesPlanetaryClimate2010}). This $f_{\sf freeze}$ coincides approximately with the outer edge of the habitable zone, defined where  CO$_2$ can no longer provide greenhouse warming because it condenses out of the atmosphere \citep{kopparapuHabitableZonesMainsequence2013}. While worlds cooler than $T_{\sf freeze}$ can obviously still have atmospheres, the sources and sinks for those atmospheres (mostly H$_2$, CH$_4$, N$_2$ in the Solar System) may differ from temperate terrestrial planets' CO$_2$-dominated atmospheres. We will include planets cooler than $T_{\sf freeze}$ in our main shoreline estimate, but we will also test the effect of restricting to only planets warm enough for CO$_2$ to remain gaseous.

With these two thresholds, we define three samples of planets. We will compare the shorelines inferred from each sample below. 
\begin{itemize}
    \item {\em No magma ocean:} Our main analysis includes all planets $f < f_{\sf magma}$, comprising 869 $A_{\sf i} = 1$ and 48 $A_{\sf i} = 0$ worlds. This sample includes the entire Solar System and all but a few of the hottest exoplanets.
    \item {\em No magma ocean, no CO$_2$ freezeout:} As an auxiliary test, we consider only planets between $f_{\sf freeze} < f < f_{\sf magma}$, including 858 $A_{\sf i} = 1$ and 17 $A_{\sf i} = 0$. This sample excludes the  outer Solar System, effectively examining the shoreline for non-molten planets out to Mars-like orbits.
    \item {\em Any temperature:} As another auxiliary test, we include all objects, with no limits on $f$, with 1026 $A_{\sf i} = 1$  and 48 $A_{\sf i} = 0$. This sample includes the ultrahot planets TOI-561 b and 55 Cnc e, which are above the $f_{\sf magma}$ flux limit.
\end{itemize}
}

\section{Fitting a Cosmic Shoreline}
\label{s:fitting} 

We construct a generative model that tries to explain the atmosphere labels $A_{\sf i}$ for planet $i$ using the predictors $f_{\sf i}$, $v_{\sf esc, i}$, $L_{\sf \star, i}$. We first define a cosmic shoreline flux $f_{\sf shoreline}$ for escape velocity $v_{\sf esc}$ and stellar luminosity $L_\star$ with the power law expression

\begin{equation}
\label{e:f_shoreline}
f_{\sf shoreline} = f_{\sf 0} \left(\frac{v_{\sf esc}}{v_{\sf esc, \oplus}}\right)^p\left(\frac{L_\star}{L_\sun}\right)^q
\end{equation}
where $f_{\sf 0}$, $p$, and $q$ are model parameters, $v_{\sf esc, \earth} = 11.18$ km/s is Earth's escape velocity, and $L_\sun = 3.828 \times 10^{26}$ W is the Sun's luminosity. We compare all fluxes to Earth's average bolometric flux $f_\earth = L_\sun/(4\pi a)^2 = 1361 \mathrm{W/m^2}$. This power law log transforms to a linear plane in 3D space

\begin{widetext}
\begin{equation}
\label{e:log_f_shoreline}
 \log_{10} \left(\frac{f_{\sf shoreline}}{f_\earth}\right) =  \log_{10} \left(\frac{f_{\sf 0}}{f_\earth}\right)  + p \cdot  \log_{10} \left(\frac{v_{\sf esc}}{v_{\sf esc, \oplus}}\right) + q \cdot  \log_{10} \left(\frac{L_\star}{L_\sun}\right).
\end{equation}
\end{widetext}

We define a distance from this shoreline in log-flux as 
\begin{widetext}
\begin{equation}
\label{e:delta}
\Delta =  \log_{10} \left(\frac{f_{\sf }}{f_\earth}\right) - \log_{10}\left(\frac{f_{\sf shoreline}}{f_\earth}\right) = \log_{10}\left(\frac{f}{f_{\sf shoreline}}\right)
\end{equation}
\end{widetext}
which is similar to the Atmosphere Retention Metric from \citet{passRecedingCosmicShoreline2025} \added{and \citet{2025_meni-gallardo_EmpiricalDeterminationCosmic}}. We use this distance to describe the probability of each planet having an atmosphere with the logistic function \citep[see][]{ivezicStatisticsDataMining2020} as
\begin{equation}
p_{\sf i} = P(A_{\sf i} = 1 | \mathbf{x}_{\sf i}, \boldsymbol{\theta} ) = \frac{1}{1+e^{\Delta_{\sf i}/w}}
\label{e:p_i}
\end{equation}
where the predictors for each datum are $\mathbf{x}_{\sf i} =$ [$\log_{10}(f_{\sf i}/f_\earth)$, $\log_{10} (v_{\sf esc, i}/v_{\sf esc, \earth})$, $\log_{10} (L_{\sf \star, i}/L_\sun)$] and the model parameters are $\boldsymbol{\theta} = [f_{\sf 0}, p, q, w]$. This logistic function smoothly transitions from 1 when $f$ is below the shoreline to 0 above, with the width parameter $w$ describing the fuzziness of the shoreline, how quickly in $\log_{10} (f/f_\earth)$ planets change from mostly having atmospheres to mostly not. The likelihood of the data ensemble $\mathbf{A}$ can be calculated by multiplying $p_{\sf i}^{A_{\sf i}}$ (= how well did we predict the presence of an atmosphere) by $(1-p_{\sf i})^{1-A_{\sf i}}$ (= how well did we predict the absence of an atmosphere) across all data points (a Bernoulli distribution):

\begin{equation}
P(\mathbf{A} | \boldsymbol{\theta}) = \prod_{i=1}^{N} p_{\sf i}^{A_{\sf i}} \cdot (1-p_{\sf i})^{1-A_{\sf i}}
\end{equation} 
This likelihood $P(\mathbf{A} | \boldsymbol{\theta})$
 and a prior $P(\boldsymbol{\theta})$ together determine the posterior probability $P(\boldsymbol{\theta} | \mathbf{A}) = P(\mathbf{A} | \boldsymbol{\theta}) P(\boldsymbol{\theta})$. For $f_{\sf 0}$, we adopt an uninformative uniform prior on $\log_{10} (f_{\sf 0}/f_\earth)$. For $p$, and $q$, we adopt the uninformative prior $P(\boldsymbol{\theta}) \propto (1+p^2)^{-3/2}\cdot (1+q^2)^{-3/2}$, which avoids the infinitely growing prior space toward high slope values and effectively represents a uniform prior on the rotation angle of the slopes in log space \citep[see][]{vanderplasFrequentismBayesianismPythondriven2014a}. For $w$, we adopt a uniform prior of $-6 > \ln w > 2$,  spanning $0.0024~\mathrm{dex} < w < 7.4~\mathrm{dex}$, or from a 0.57\% change in $f$ at the narrowest to a factor of $2.5\times10^7$ change in $f$ at the widest; practically we find that allowing narrower widths than this can reveal sharp discontinuities that become difficult to sample. 

To account for measurement uncertainties on the predictors, the expression for $p_{\sf i}$ in Equation \ref{e:p_i} should be marginalized over the distribution of true (unknown) values for each planet's $\mathbf{x}_{\sf i}$. While this marginalization can sometimes be done analytically (as in the mass-radius fit in \S\ref{f:mass-radius}), we were unable to find a simple analytic expression and instead relied on the {\em remarkable} efficiency of \texttt{numpyro}'s NUTS sampler to do this marginalization numerically. We introduced three parameters for {\em each} datapoint (\added{up to 1,074 planets or 3,222 parameters}) to represent the true values of $\mathbf{x}_{\sf i}$, with normal priors centered on the measured values and the uncertainties as their widths, and we sampled these alongside the 4 parameters $\boldsymbol{\theta}$ we actually care about. 

We used \texttt{numpyro} with NUTS to sample from this posterior, running 4 chains with 5,000 warm-up steps and 50,000 samples each, always achieving a Gelman-Rubin statistic of 1.0 across the chains and usually achieving an effective sample size always (and sometimes much) larger than 1,000 (\href{https://github.com/zkbt/shoreline/blob/main/notebooks/3-fit-one-shoreline.ipynb}{\texttt{</>}}). Even with the thousands of hyperparameters we use to marginalize over measurement uncertainties, this sampling takes only a few minutes on a modern MacBook Pro. We repeat these fits, with and without uncertainties, across various subsamples of the data (\href{https://github.com/zkbt/shoreline/blob/main/notebooks/4-fit-many-shorelines.ipynb}{\texttt{</>}}).

\begin{figure*}[ht!]
\includegraphics[width=\textwidth]{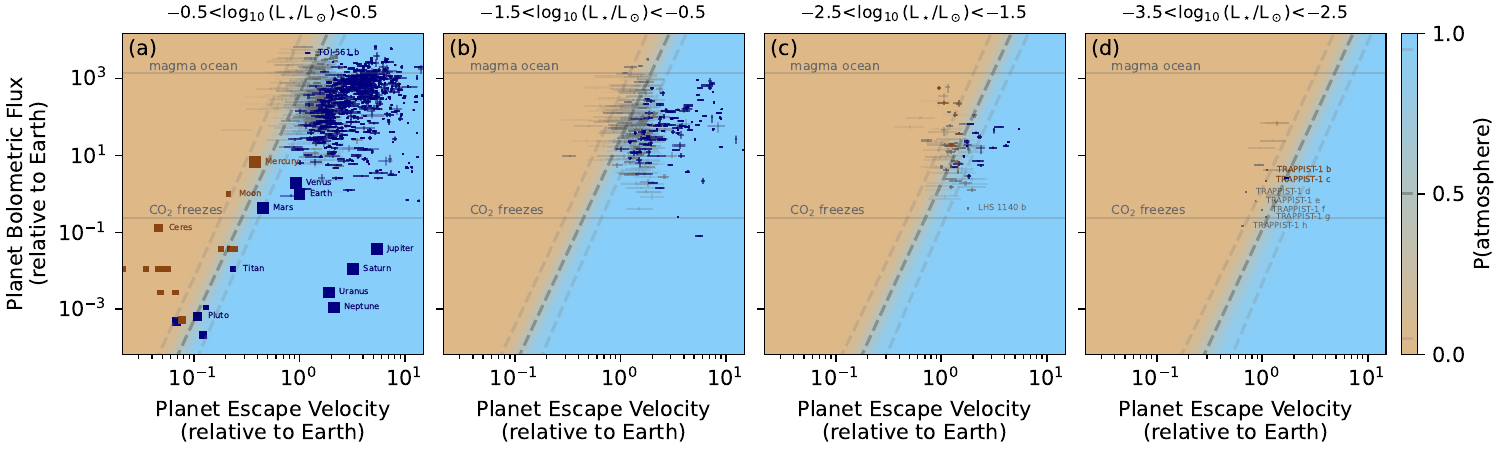}
\includegraphics[width=\textwidth]{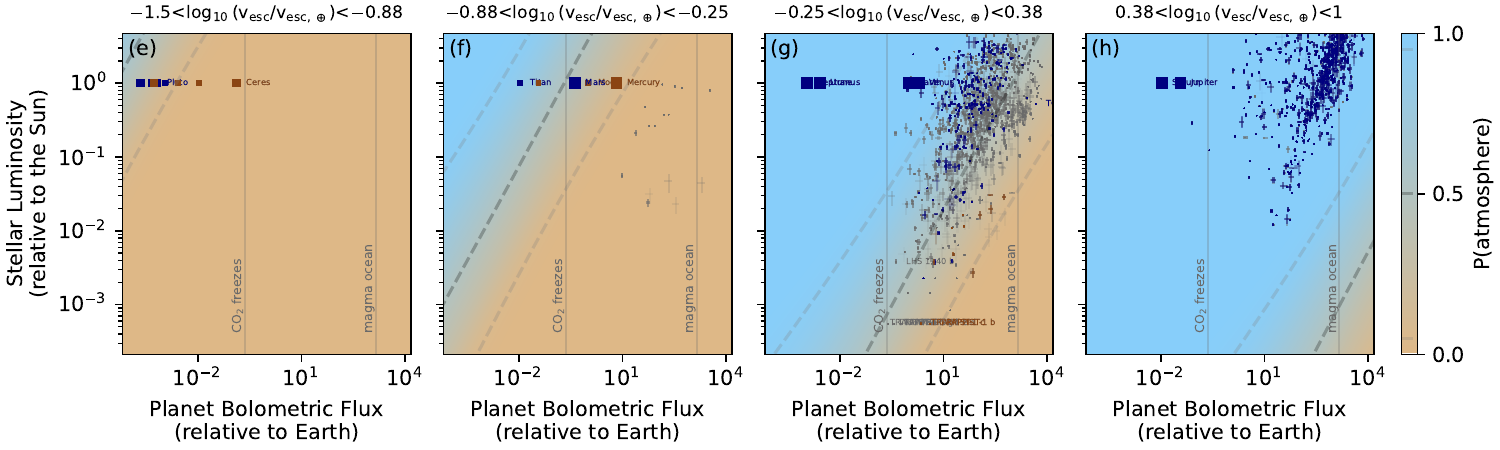}
\includegraphics[width=\textwidth]{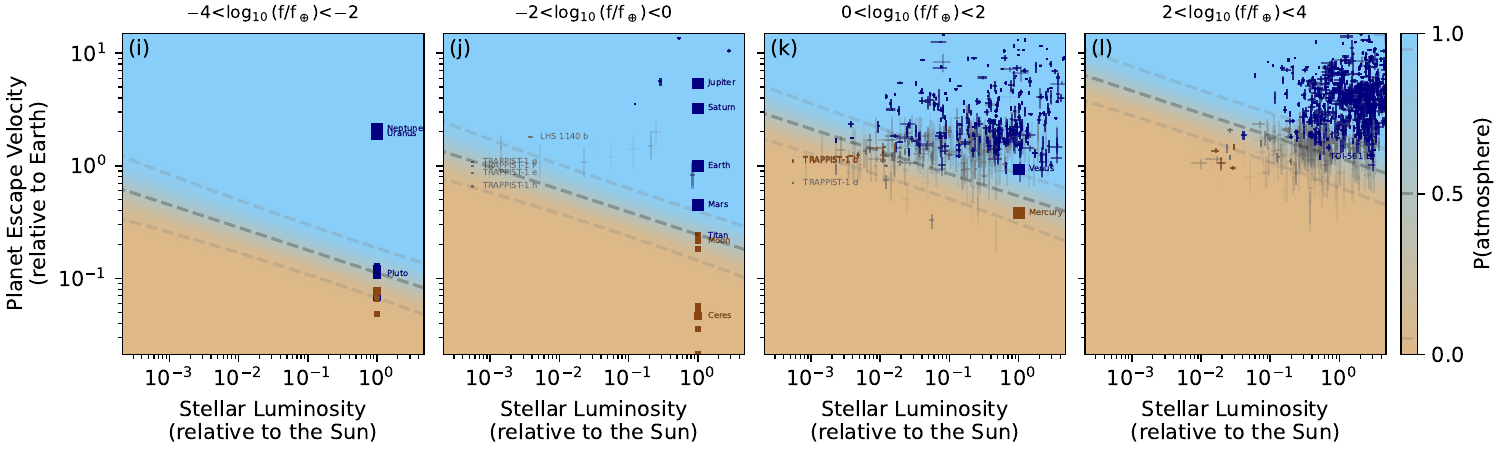}

\caption{A cosmic shoreline dividing exoplanets (errorbars) and Solar System planets (squares) with any type of atmosphere or global surface volatiles (blue symbols, $A_{\sf{i}}=1$) from those without (brown symbols, $A_{\sf{i}}=0$). \added{Magma ocean planets ($T_{\sf eq} > 1700$K) were excluded from the fit but are still shown for context, as are planets without definitive atmosphere constraints  (gray symbols, $A_{\sf{i}}=?$).} The shoreline defines a plane in the 3D space of ($f$, $v_{\sf esc}$, $L$); each row shows slices that consider a narrow range of stellar luminosity (top), planet escape velocity (middle), and planet flux (bottom). Background colors indicate the modeled probability of an atmosphere at each location (sandy brown for $p_{\sf i} = 0$, water blue for $p_{\sf i}$ = 1), marginalizing over the parameter uncertainties and the width of the slice; contours (dashed lines) highlight atmosphere probabilities of 5\%, 50\%, 95\% (\href{https://github.com/zkbt/shoreline/blob/main/notebooks/6-plot-shorelines-on-their-own.ipynb}{\texttt{</>}}).}
\label{f:shoreline}
\end{figure*}

\begin{figure}
\begin{interactive}{animation}{figures/shoreline-vfL+few-annotations-animated.mp4}
\includegraphics[width=\columnwidth]{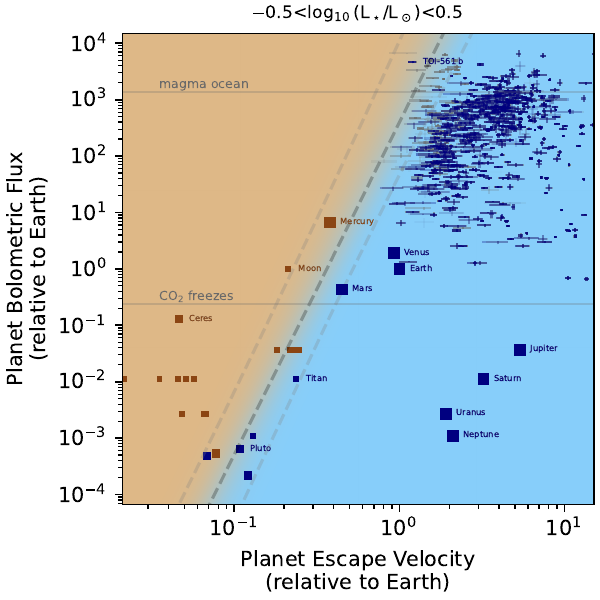}

\end{interactive}
\caption{The same cosmic shoreline as the top row of Figure \ref{f:shoreline}, but expressed as an animation that is available in the HTML version of the article, showing the 3D shape of the cosmic shoreline by stepping through slices of changing stellar luminosity (\href{https://github.com/zkbt/shoreline/blob/main/notebooks/6-plot-shorelines-on-their-own.ipynb}{\texttt{</>}}).
}
\label{f:shoreline-animated}

\end{figure}

\section{Shorelines in 3D}
\label{s:shorelines}

\added{Below, we present this newly inferred shoreline model and explore the meaning of its parameters in \S\ref{s:shoreline-no-magma}. We present additional tests that probe the sensitivity of this shoreline to changing planet samples or modeling assumptions in \S\ref{s:shoreline-no-freeze},  \S\ref{s:shoreline-no-giants},  \S\ref{s:shoreline-no-uncertainty}.
}

\subsection{A New Cosmic Shoreline}
\label{s:shoreline-no-magma}

\added{Figures \ref{f:shoreline} and \ref{f:shoreline-animated} show the inferred shoreline model. All Solar System worlds and exoplanets in the ``no magma ocean'' sample were used to determine the shoreline parameters. Although not included in the fit,  the ultrahot magma- ocean planets are still shown for context.} Because the probability of an atmosphere $A$ is a function in a 3D volume, it can be a little tricky to visualize in a single plot. In Figure \ref{f:shoreline} we display this 3D volume in slices: each row holds one dimension fixed to a narrow range and visualizes the other two dimensions on the $x$ and $y$ axes, and each column displays a different range of values for the fixed dimension. The background color shows the modeled probability of an atmosphere at each $(x, y)$ location in each slice, marginalized (= integrated, see \citealt{hoggDataAnalysisRecipes2010a, siviaDataAnalysisBayesian2011, vanderplasFrequentismBayesianismPythondriven2014a, ivezicStatisticsDataMining2020}) over both the width of the slice and the uncertainties on the model parameters. Even for infinitely thin slices with no parameter uncertainties the transition would still appear fuzzy due to the intrinsic width $w$ (see Equation \ref{e:p_i}). \added{Figure \ref{f:shoreline-animated} shows an animated version, stepping through slices in stellar luminosity. }

The top row (a-d) of Figure \ref{f:shoreline} holds $L_\star$ as the fixed dimension, decreasing from solar type host stars on the left to the latest possible M dwarf stars on the right. The centers of these luminosity ranges (1$L_\sun$, 0.1$L_\sun$, 0.01$L_\sun$, 0.001$L_\sun$) correspond to main-sequence spectral types (G2, K7, M3.5, and M6) according to the \citet{pecautINTRINSICCOLORSTEMPERATURES2013} sequence. Panel (a) shows $v_{\sf esc}$ and bolometric flux $f$ for both Solar System objects (all with $L = 1.0L_\sun$) and exoplanets with host stars within a factor of $\sqrt{10}$ of the Sun's luminosity; it is the closest analog to Figure 1 from ZC17, which shows similar quantities but without the restriction on exoplanet host star type. The shoreline in this row has a slope of $p$ (see Equation \ref{e:log_f_shoreline}) and reading from left to right appears to recede \citep[to borrow the visual metaphor from][]{passRecedingCosmicShoreline2025} down and to the right, with the bolometric threshold $f_{\sf shoreline}$ decreasing at fixed $v_{\sf esc}$ toward lower luminosity stars. 

The middle row (e-h) shows shoreline slices for different fixed $v_{\sf esc}$, increasing from tiny low-mass dwarf planets on the left to gas giants on the right. Only Solar System objects are known at low $v_{\sf esc}$ (e), but for Earth-like $v_{\sf esc}$ values (g) exoplanet atmosphere data become available either as radii large enough to require volatiles or as rocky planets with JWST hot dayside brightness temperatures disfavoring thick atmospheres. In these slices the visible slope is $1/q$, with cooler less luminous stars having lower maximum allowable flux levels $f_{\sf shoreline}$ for atmospheres to survive. 

The bottom row (i-l) shows the shoreline for different fixed $f$, increasing from the cold outer regions of the Solar System on the left to the very hottest exoplanets on the right. The slope of the shoreline in this projection is $-q/p$ and indicates a larger $v_{\sf esc}$ is necessary in order for lower $L_\star$ hosts to permit atmospheres. For temperate planets (j), if we imagine shrinking the host star luminosity while keeping $f$ constant at $f_{\oplus}$, Mars-sized planets would be unable to retain atmospheres around stars less luminous than $0.1~L_\sun$, and Earth/Venus-size planets would likely lose atmospheres somewhere between $10^{-3} - 10^{-2}~L_\sun$.

\begin{figure}[ht!]
\includegraphics[width=\columnwidth]{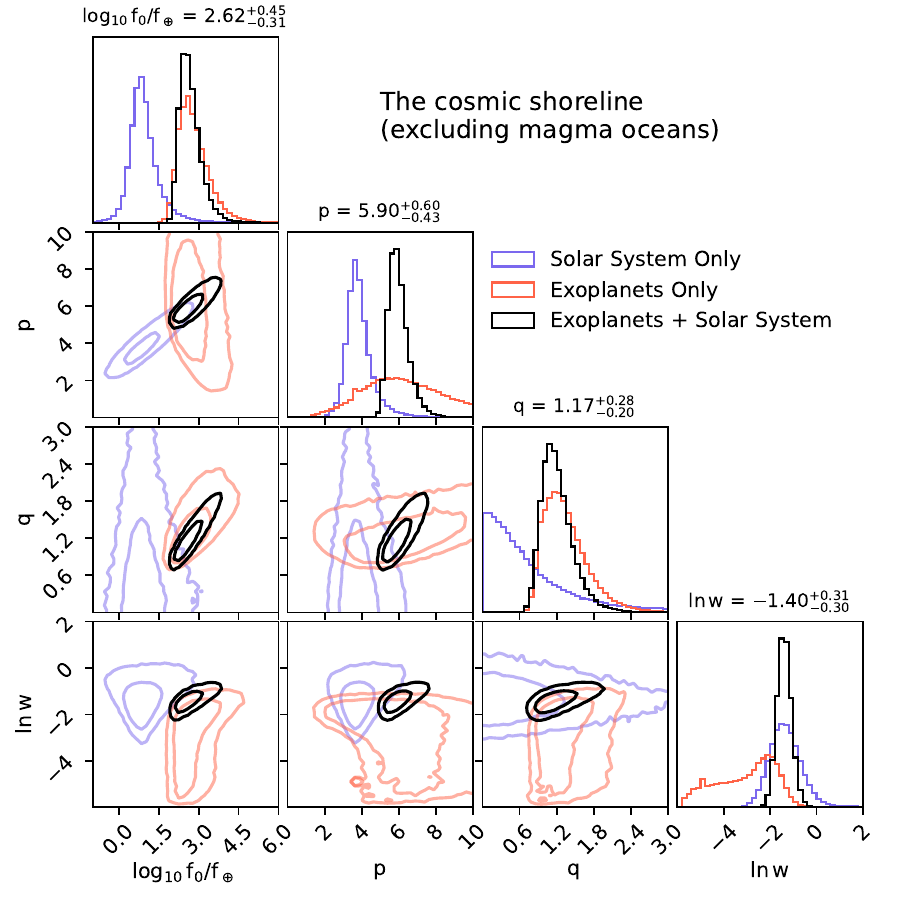}
\caption{Cosmic shoreline parameter posterior probabilities, including only planets cool enough to avoid global magma oceans. These parameters define the atmosphere probability shown in Figure \ref{f:shoreline}. Panels show marginalized 1D histograms (diagonal) and marginalized 2D distributions (off-diagonal) with contours that enclose 68.3\% and 95.4\% probability. Titles along the diagonal show central $68.3\%$ confidence intervals for the exoplanets + Solar System joint fit. The model parameters define a shoreline via $\log_{10} (f_{\rm shoreline}/f_\earth) = \allowbreak \log_{10} (f_{\sf 0}/f_\earth) + \allowbreak p \log_{10} (v_{\sf esc}/v_{\sf esc, \earth}) + \allowbreak q \log_{10} (L_\star/L_\sun)$, with $w$ representing the logistic width parameter setting the fuzziness of the shoreline  (\href{https://github.com/zkbt/shoreline/blob/main/notebooks/5-print-and-visualize-posteriors.ipynb}{\texttt{</>}}).}
\label{f:posteriors-no-magma}
\end{figure}

Figure \ref{f:posteriors-no-magma} shows the posterior probability distributions for the shoreline parameters. We show both the main fit including all planets together and what we might learn from just Solar System or just exoplanets each by themselves. We compare these posteriors with \texttt{corner.py} \citep{foreman-mackeyCornerpyScatterplotMatrices2016}, with contours in each 2D panel that enclose 68.3\% and 95.4\% of the probability marginalized over other parameters. 

We find the intercept to be $\log_{10}{f_{\sf 0}} = \allnomagmalogfo$, meaning that an Earth-size planet orbiting a Sun-like star should on average be able to retain an atmosphere with bolometric flux levels $f/f_\earth$ up to about $10^{\allnomagmalogfojustvalue} = \allnomagmafojustvalue$. If we moved Earth inward toward the Sun, hydrogen and the hope of habitability would be lost long before this limit, likely leaving heavily oxidized CO$_2$/O$_2$ atmospheres. \added{Yet, these dense atmospheres apparently might be retained even up to surprisingly high temperatures.}

The escape velocity slope $p = \allnomagmap$ is steeper than $p=4$ chosen by ZC17 and means that a factor of $10\times$ increase in $v_{\sf esc}$ causes the critical flux $f_{\sf shoreline}$ to move up by about $10^{\allnomagmapjustvalue}$. Notably, for Solar System planets alone we find $p = \solarnomagmap$ and for exoplanets alone we find $p = \exonomagmap$. Each is individually consistent with $p=4$, so the higher joint slope reflects a compromise where these two different samples that mostly occupy different regions of parameter space can agree (see Figure \ref{f:posteriors-no-magma}).

The stellar luminosity slope $q = \allnomagmaq$ means that if we decrease the luminosity of the host star by $10\times$, the maximum flux that permits atmospheres $f_{\sf shoreline}$ decreases by a factor $10^{\allnomagmaqjustvalue}$. This is steep! If we take $f_{\sf hz} = f_\earth$ as a crude approximation for the habitable zone (neglecting the important dependence on the stellar spectrum; \citealt{kopparapuHabitableZonesMainsequence2013}), this fit implies $f_{\sf shoreline} < f_{\sf hz}$ for stars less luminous than $\log_{10} (L_\star/L_\sun) = \allnomagmalogLnohz$ or $L_\star/L_\sun = \allnomagmaLnohz$, corresponding to M4V spectral type \citep{pecautINTRINSICCOLORSTEMPERATURES2013} or  $0.25 \mathrm{M_\sun}$ mass \citep{pinedaMdwarfUltravioletSpectroscopic2021}. \added{We further discuss the steep $q$ slope's implications for habitability in \S\ref{s:implication-hz}.}

The intrinsic width of the shoreline $\ln w = \allnomagmalnw$, or $w = \allnomagmaw~\mathrm{dex}$, means than if $f$ increases by a factor of $10^{\allnomagmawjustvalue}$ above $f_{\sf shoreline}$ then the probability of an atmosphere drops from $50\%$ to $1/(1 + e^{1}) = 27\%$ (Equation \ref{e:p_i}). \added{To translate into more familiar probabilities from a normal distribution, the chance of having an atmosphere drops to 32\% (100\% - 68\%, which might colloquially be called a 1$\sigma$ upper limit) and 5\% (2$\sigma$ upper limit) at $0.75w$ and $2.94w$, respectively. We define $w_{95}$ as the width of the transition from atmospheres being 95\% likely to only 5\% likely, which  corresponds to $w_{95} = 5.89w = \allnomagmawninetyfive~\mathrm{dex}$ or a factor of $\allnomagmawninetyfiveasfluxfactor \times$ in flux (\href{https://github.com/zkbt/shoreline/blob/main/notebooks/x-plot-logistic-probabilities.ipynb}{\texttt{</>}}). The blending of colors and the dashed $5-95\%$ lines in Figures \ref{f:shoreline} and \ref{f:shoreline-animated} include the fuzziness from this $w$, along with marginalization over the uncertain shoreline parameters and the finite width of the visualization slice. }

\added{
\subsection{Sensitivity to planet flux limits.}
\label{s:shoreline-no-freeze}

Our main shoreline fit uses planets from the ``no magma ocean'' sample (\S\ref{s:flux-limits}), excluding planets that receive fluxes above the magma ocean flux limit $f_{\sf magma}$ to avoid the larger volatile inventories accessible in these planets' atmospheres. Here, we briefly explore the effect of applying different flux limits to the sample of planets we consider.

\begin{figure*}[ht!]
\includegraphics[width=\textwidth]{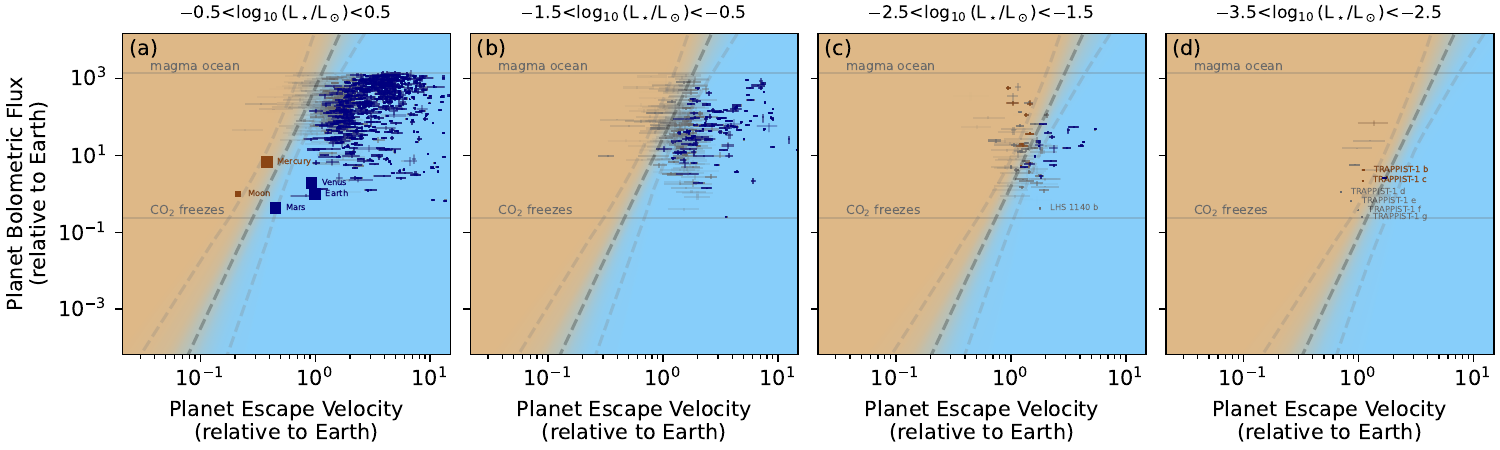}
\caption{A cosmic shoreline exactly as in the top row of Figure \ref{f:shoreline}, but specifically targeting $\rm{CO}_2$-dominated atmospheres in the ``no magma ocean, no CO$_2$ freezeout" sample with equilibrium temperatures warm enough for $\rm{CO_2}$ to exist as a gas (${\rm T_{CO_2}} > 194 \mathrm{K}$) and cool enough that a global magma ocean is less likely (${\rm T_{magma}} < 1700 \mathrm{K}$). This more narrowly defined cosmic shoreline matters because $\rm CO_2$ is both likely in warm secondary atmospheres and relatively easier to detect for exoplanets in thermal emission with JWST (\href{https://github.com/zkbt/shoreline/blob/main/notebooks/6-plot-shorelines-on-their-own.ipynb}{\texttt{</>}}).}
\label{f:shoreline-CO2}
\end{figure*}

If we zoom in, also excluding planets with fluxes below the CO$_2$ freeze-out flux limit $f_{\sf freeze}$ (``no magma ocean, no CO$_2$ freezeout''), we find the shoreline fit shown in Figure \ref{f:shoreline-CO2}. The advantage of removing the colder worlds is that we can more confidently compare broadly similar atmospheres, where planetary climates are fundamentally set by the planets' resilience for retaining gaseous CO$_2$. Much colder CH$_4$ and N$_2$ atmospheres or surface volatile reservoirs like Titan's or Pluto's may represent fundamentally different formation and evolution pathways, and might therefore exhibit different shorelines. The disadvantage of excluding cool planets is that we remove the entire outer Solar System, which provides considerable leverage on the shoreline slopes (particularly $p$). The inferred slopes $p = \allnomagmanofreezep$ and $q=\allnomagmanofreezeq$ (Figure \ref{f:posteriors-kinds}) indeed do have larger uncertainties than for the main sample. Yet, the slopes are consistent between the two samples. 

\begin{figure}[ht!]
\includegraphics[width=\columnwidth]{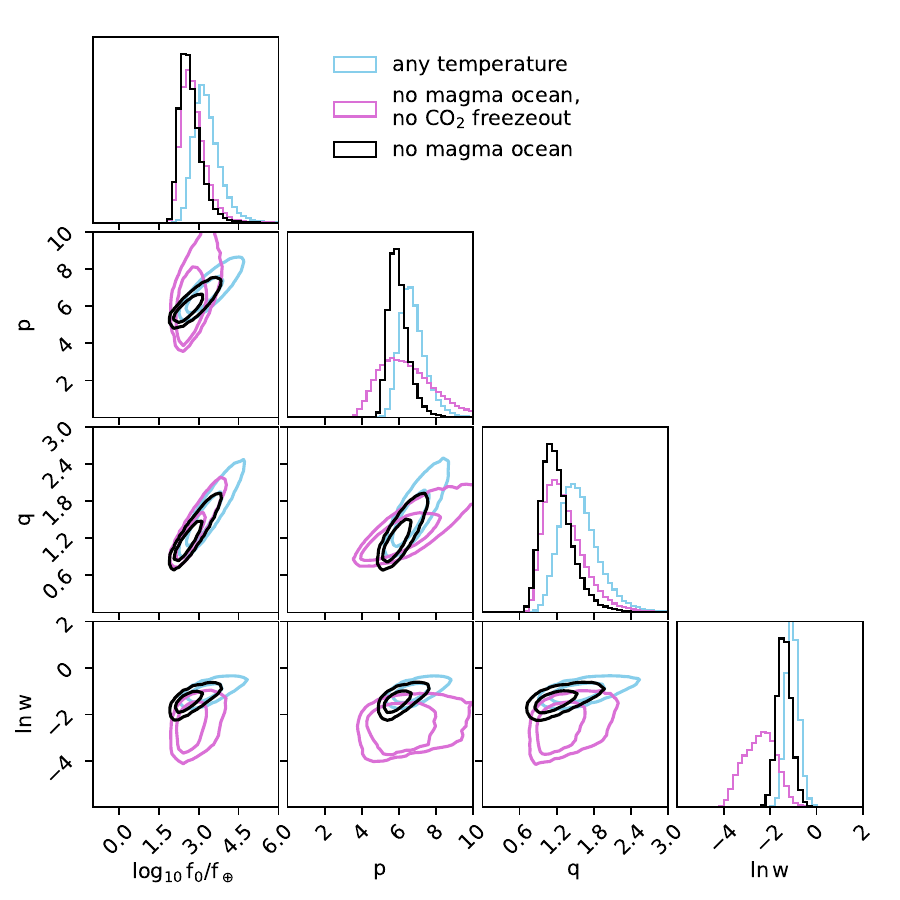}

\caption{Cosmic shoreline parameter posterior probabilities considering planets that fall within three different bolometric flux limits, using exoplanet and Solar System data in all fits. Panels show marginalized 1D histograms (diagonal) and marginalized 2D distributions (off-diagonal) with contours that enclose 68.3\% and 95.4\% probability. The model parameters define a shoreline via $\log_{10} (f_{\rm shoreline}/f_\earth) = \allowbreak\log_{10} (f_{\sf 0}/f_\earth) +\allowbreak p \log_{10} (v_{\sf esc}/v_{\sf esc, \earth}) + \allowbreak q \log_{10} (L_\star/L_\sun)$, with $w$ representing the logistic width parameter setting the fuzziness of the shoreline (\href{https://github.com/zkbt/shoreline/blob/main/notebooks/5-print-and-visualize-posteriors.ipynb}{\texttt{</>}}).}
\label{f:posteriors-kinds}
\end{figure}

If we zoom out, excluding no planets, even those hot enough for continuously molten surfaces (``any temperature''), the shoreline shifts. One planet, TOI-561 b \citep[][shown in Figure \ref{f:shoreline}a]{2025_teske_ThickVolatileAtmosphere}, nudges the shoreline toward steeper slopes $p=\allanyp$ and $q=\allanyq$ (Figure \ref{f:posteriors-kinds}) to try to capture the apparent atmosphere on this ultrahot world transiting a roughly Sun-like star. That TOI-561b prefers a different shoreline than the ``no magma ocean'' sample marginally confirms the hypothesis that lava worlds may have access to deeper pools atmospheric volatiles and that their abilities to support atmospheres cannot be simply extrapolated to more temperate planets.

Curiously, the width of the shoreline shows a tentative trend of decreasing $w$ toward tighter population flux limits (Figure \ref{f:posteriors-kinds}) with $\ln w = \allanylnw$ for ``any temperature'', $\ln w = \allnomagmalnw$ for ``no magma ocean'', $\ln w = \allnomagmanofreezelnw$ for ``no magma ocean, no CO$_2$ freezeout''. One possibility is that the width $w$ on broad global scales is partially capturing unknown curvature or topography along the shoreline. A lower $w$ in the zoomed-in sample might hint at the log-linear shoreline being a better local description on increasingly fine scales. More thoroughly testing for such local geography to the shape of the shoreline will require many more planets with reliable atmosphere labels. }

\begin{figure}[ht!]
\includegraphics[width=\columnwidth]{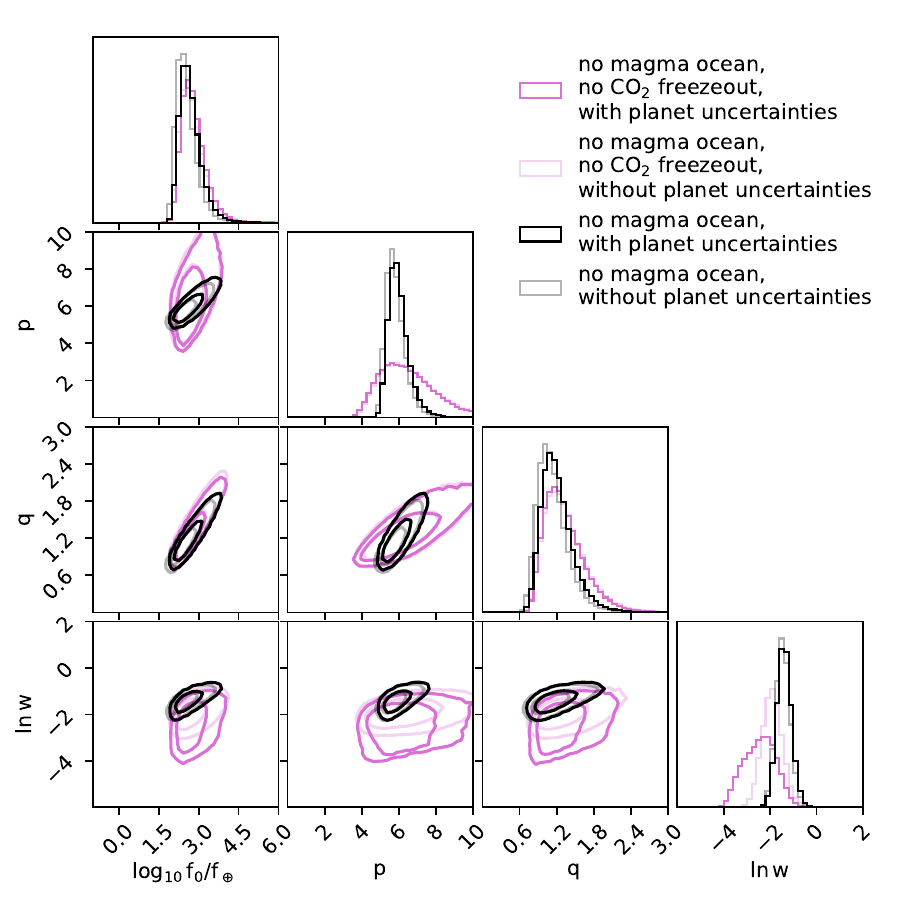}
\caption{Posteriors inferred with (intense lines) and without (faint lines) accounting for uncertainties on planet properties ($f$, $v_{\sf esc}$, $L_\star$). Two planet flux limits are shown, using exoplanet and Solar System data in all fits. Marginalizing over uncertainties barely matters for the large ``no magma ocean'' sample but broadens the possible intrinsic shoreline widths $w$ for the smaller ``no magma ocean, no CO$_2$ freezeout'' sample to lower values, because $w$ is less necessary when planets' scatter can be explained by their own measurement uncertainties (\href{https://github.com/zkbt/shoreline/blob/main/notebooks/5-print-and-visualize-posteriors.ipynb}{\texttt{</>}}).}
\label{f:posteriors-uncertainties}
\end{figure}

\subsection{Sensitivity to Including Gas Giant Planets}
\label{s:shoreline-no-giants}

In the above fits, we did not place upper limits on planet escape velocity or radius, allowing all giant planets to participate in sculpting the shoreline, as in ZC17. Hot Jupiters could potentially bias the shoreline slope, in that they might lose tens of Earth masses of atmosphere but still appear as $A_{\sf i} = 1$. We tested the sensitivity of the inferred shoreline to this concern by repeating the fits including only planets smaller than Neptune. \added{We found no significant changes to the parameters, likely because hot Jupiters all have such high surface gravities that they fall well within the volume where atmospheres are inevitably strongly predicted.}

\subsection{Sensitivity to Planet Parameter Uncertainties}
\label{s:shoreline-no-uncertainty}
We test the impact that measurement uncertainties on the predictors $v_{\sf esc}, f, L_\star$ have the inferred shoreline. Figure \ref{f:posteriors-uncertainties} shows parameter posteriors with and without including parameter uncertainties for two of the planet subsamples. Including planet uncertainties slightly broadens distributions and also allows for lower values of $w$, where disagreeing labels at fixed shoreline distance $\Delta_{\sf i}$ can be a little more explained by uncertainties on the predictors and thus requiring less intrinsic fuzziness. This is especially true for the smallest ``no magma ocean, no CO$_2$ freezeout'' sample; with fewer planets, the uncertainty on each planet's position relative to the shoreline matters more. Still, the differences are minor for all fits, likely because parameter uncertainties for most well-characterized planets are smaller than the fuzziness already being captured in the intrinsic width $w$. 

\section{Physical Interpretation}
\label{s:physics}

Atmospheric loss is fundamentally a matter of energy balance: whatever energy a planet receives from its star (or still-cooling interior; see \citealt{guptaSculptingValleyRadius2019}), must either radiate away or be carried away in the gravitational potential energy of escaping gas \citep{lewisPlanetsTheirAtmospheres1984, chamberlainTheoryPlanetaryAtmospheres1987}. The incoming energy need not be radiative, with particles and fields in the stellar wind also driving loss, and moreso during coronal mass ejections \citep{lammerCoronalMassEjection2007, jakoskyMAVENObservationsResponse2015}. Modeling efforts beyond ZC17 have included various atmospheric sources and sinks to understand where atmospheres can or cannot flourish \citep{tianTHERMALESCAPESUPER2009,
lugerHabitableEvaporatedCores2015, 
owenEvaporationValleyKepler2017, 
wyattSusceptibilityPlanetaryAtmospheres2020, 
guptaCaughtActCorepowered2021,
chatterjeeNovelPhysicsEscaping2024, 
chinRolePlanetaryRadius2024a, 
giallucaImplicationsThermalHydrodynamic2024,
teixeiraCarbondeficientEvolutionTRAPPIST1c2024,
zengCosmicHydrogenIce2024,
vanlooverenAiryWorldsBarren2024,
vanlooverenHabitableZoneAtmosphere2025, 
leeCarvingEdgesRocky2025,
jiCosmicShorelineRevisited2025}. Here, we briefly try to contextualize the newly inferred shoreline parameters through the lens of hydrodynamic escape, as the most efficient path for atmospheric erosion in extreme environments where XUV heating can overwhelm infrared cooling in the tenuous upper atmosphere and drive fluid flows \citep{sekiyaDissipationRareGases1980, watsonDynamicsRapidlyEscaping1981}. 

For the flux slope $p$, we can consider a simplified model of energy-limited escape, where some fraction $\epsilon_{\sf esc}$ of incoming power from $f_{\sf XUV}$ radiation converts directly into gravitational potential energy of outflowing atmosphere \citep{watsonDynamicsRapidlyEscaping1981}, can be written as $\epsilon_{\sf esc} f_{\sf XUV} \pi R_{\sf XUV}^2 \approx GM \dot{M}_{\sf atm}/R_{\sf atm}$ with $\pi R_{\sf XUV}^2$ as the planet's cross-section to high-energy radiation, $\dot{M}_{\sf atm}$ as the atmospheric mass loss rate, and $R_{\sf atm}$ as the effective radius from which atmosphere is escaping. If we neglect important radiative and tidal effects \citep[see][]{lammerAtmosphericLossExoplanets2003, erkaevRocheLobeEffects2007}, crudely approximate $R_{\sf XUV} \approx R_{\sf atm} \approx R$, parameterize the high-energy flux as some fraction of the bolometric flux $f_{\sf XUV} = \epsilon_{\sf XUV} f$,  imagine the atmospheric volatile budget eroded over the system age $t$ to be some fraction of planet mass $\dot{M}_{\sf atm} = \epsilon_{\sf atm} M/t$, and define $\epsilon_{\sf ?} = \epsilon_{\sf atm} \cdot \epsilon_{\sf esc}^{-1} \cdot \epsilon_{\sf XUV}^{-1}$ as a {\em deeply} uncertain combined efficiency factor, we would find a shoreline that scales with bolometric flux as $f \propto \epsilon_{\sf ?}  M^2/R^3  \propto \epsilon_{\sf ?}  v_{\sf esc}^4/R  \propto \epsilon_{\sf ?}  v_{\sf esc}^3\sqrt{\rho}$ as in Figure 3 of ZC17. If we use the mass-radius relation in Figure \ref{f:mass-radius} to estimate $R\propto v_{\sf esc}^{\mrpowerRvesc}$ or $M \propto v_{\sf esc}^{\mrpowerMvesc}$ for rocky planets, we find $f \propto  \epsilon_{\sf ?}  v_{\sf esc}^p$ with $p \approx \mrpowerpjustvalue$. Strong dependencies lurk inside $\epsilon_{\sf ?}$ that could tilt the flux slope $p$ away from this cartoon $p=\mrpowerpjustvalue$ value, either globally or locally: $\epsilon_{\sf atm}$ depends on volatile delivery history, interior-atmosphere exchange, instellation, and tides \citep{elkins-tantonRangesAtmosphericMass2008, schaeferPredictionsAtmosphericComposition2016c, kiteAtmosphereinteriorExchangeHot2016, seligmanPotentialMeltingExtrasolar2024}, and $\epsilon_{\sf esc}$ depends (at least) on instellation, planet mass, and composition \citep{murray-clayAtmosphericEscapeHot2009, owenPlanetaryEvaporationUV2012, owenEvaporationValleyKepler2017, chatterjeeNovelPhysicsEscaping2024, jiCosmicShorelineRevisited2025, leeCarvingEdgesRocky2025}.

Another useful reference slope $p$ comes from a common threshold for mass loss: the escape parameter $\lambda =  v_{\sf esc}^2/v_{\sf thermal}^2$, where $v_{\sf thermal} = \sqrt{2k_{\sf B} T/m}$  is the thermal speed of the gas, with $k_{\sf B}$ as the Boltzmann constant, $T$ as temperature, and $m$ as the mass each escaping atom/molecule \citep[see][]{schallerVolatileLossRetention2007, johnsonExospheresAtmosphericEscape2008, gronoffAtmosphericEscapeProcesses2020}. If we calculate this escape parameter $\lambda$ with planets' zero-albedo instantaneous equilibrium temperature $T = T_{\sf eq}  \propto f^{1/4}$ (horribly inaccurately for thick atmospheres because it ignores XUV heating, but effectively setting a lower limit on atmospheric temperatures), constant values $\lambda$ would correspond to $v_{\sf esc}^2 \propto f^{1/4}$ and a shoreline slope $p=8$. One reason the slope in energy-limited escape is  shallower than this is because XUV-heated exospheres converge through infrared cooling thermostats to similar hot temperatures despite strongly varying incoming fluxes \citep{chamberlainUpperAtmospheresPlanets1962, murray-clayAtmosphericEscapeHot2009, chatterjeeNovelPhysicsEscaping2024}. That our inferred slope $p=\allnomagmap$ falls between the cartoon energy-limited and constant escape parameter limits is encouraging, but gleaning reliable insights into atmospheric evolution requires more detailed predictive modeling of the flux slope $p$. 

For the stellar luminosity slope $q$, we might interpret it as setting the fraction of light a star emits in the XUV via $\epsilon_{\sf XUV} = L_{\sf XUV}/L_{\star} = \epsilon_{\sf XUV, \sun} (L_\star/L_\sun)^{-q}$ where $\epsilon_{\sf XUV, \sun}$ is the solar XUV fraction \citep[$2\times10^{-6}$ for the quiet Sun and higher when integrated over its lifetime][]{ woodsSolarIrradianceReference2009a, franceMUSCLESTreasurySurvey2016}. Positive shoreline slopes $q > 0$ correspond to fainter stars emitting fractionally more of their luminosity in the XUV \citep[as they do;][]{wilsonMegaMUSCLESTreasurySurvey2025b}, thus requiring the threshold bolometric flux $f_{\sf shoreline}$ to decrease to keep $F_{\sf XUV}$ fixed. A single power law is clearly only an approximation to a more complicated picture: stars' XUV spectra are messy functions of age \citep{ribasEvolutionSolarActivity2005, wrightSTELLARACTIVITYROTATIONRELATIONSHIPEVOLUTION2011, pinedaFarUltravioletMdwarf2021, duvvuriHighenergySpectrumYoung2023b, kingStellarXRayVariability2025}, stellar type \citep{linskyIntrinsicExtremeUltraviolet2014, richey-yowellHAZMATUltravioletXRay2019, peacockHAZMATVIEvolution2020, wilsonMegaMUSCLESTreasurySurvey2025b}, rotational history \citep{irwinMonitorProjectRotation2007, loydHAZMATVIIEvolution2021, johnstoneActiveLivesStars2021}, and flaring activity  \citep{franceHighenergyRadiationEnvironment2020d, diamond-loweHighenergySpectrumNearby2021a, feinsteinAUMicroscopiiFarUV2022a}. ZC17 integrated old scaling relations to estimate an $F_{\sf XUV}$ scaling that translates to $q_{ZC17}=0.6$ (their Equation 26). \citet{passRecedingCosmicShoreline2025} updated this integral with modern M dwarf data, provided $F_{\sf XUV}$ in mass bins spanning $0.1-0.3 \mathrm{M_\sun}$ ($-3.1< \log_{10} L_\sun < -2.0$; their Table 1), and found the ZC17 expression under-predicted historic $F_{\sf XUV}$ fluences by $2-3\times$ for these mid-to-late M dwarfs. The \citet{passRecedingCosmicShoreline2025} estimates do not follow a constant $d\ln \epsilon_{\sf XUV}/d\ln L_\star$ across the mass range but are bounded by scalings of $q_{P25}={0.79}^{+0.04}_{-0.11}$ (median and full range, for different mass bins) relative to the Sun (\href{https://github.com/zkbt/shoreline/blob/main/notebooks/x-plot-pass-xuv-comparison.ipynb}{\texttt{</>}}). Our inferred $q = \allnomagmaq$ is higher than these estimates, suggesting an even stronger trend toward M dwarfs being inhospitable to planetary atmospheres.

\citet{vanlooverenHabitableZoneAtmosphere2025} modeled escape across stellar type including realistic stellar/rotational/activity evolution and self-consistent XUV-heated thermospheres \citep{johnstoneUpperAtmospheresTerrestrial2018}, finding thermal escape from stars' most active periods was sufficient to erode CO$_2$/N$_2$ atmospheres out to the habitable zone for all stars less massive than about $0.4 \mathrm{M_\sun}$ ($\log_{10} (L_\star/L_\sun) =-1.7$, see \citealt{pinedaMdwarfUltravioletSpectroscopic2021}). We can translate this statement about atmospheric retention in the habitable zone via $q = - \left[\log_{10} (f_{\sf 0}/f_{\sf shoreline}) + p \log_{10} (v_{\sf esc}/v_{\sf esc, \earth})\right]/\log_{10} (L_\star/L_\sun)$ with $f_{\sf shoreline} = f_{\sf hz} = f_{\earth}$ and $v_{\sf esc}/v_{\sf esc, \earth}=1$, finding $q_{VL25} = \allnomagmaqfromard$. That our inferred slope of $q = \allnomagmaq$ is consistent with this theoretical prediction suggests interpreting $q$ as primarily representing a rough XUV scaling might be reasonable. In the future, more detailed modeling of how drivers of atmospheric loss scale with stellar luminosity, including both XUV and other non-thermal drivers like stellar wind properties, could improve on the simple power law with slope $q$ assumed here.

\added{
\section{Implications}
Here, we place this new inferred shoreline in context. We compare to ZC17 in \S\ref{s:implication-zc17}, predict atmosphere probabilities for planned Rocky Worlds targets in \S\ref{s:implications-rocky-worlds}, and consider the overlap between the cosmic shoreline and the habitable zone in \S\ref{s:implication-hz}.

\subsection{Comparison to ZC17}
\label{s:implication-zc17}

How does this updated shoreline model compare to the original ZC17 formulation? Figure \ref{f:shoreline-zc17} shows how this new shoreline compares to the traditional estimate of the XUV-driven cosmic shoreline \citep[Figure 2 of ][]{zahnleCosmicShorelineEvidence2017a}. We neglect parameter uncertainties on $\log_{10}(f_0/f_\oplus), p, q, \ln w$. We collapse all stellar luminosities onto one plot, using color to indicate $L_\star$ for individual planets and the appropriate $L_\star$-dependent shoreline to which they should be compared. The slope $p=\allnomagmap$ is steeper than ZC17's $p=4$. The dependence on stellar luminosity $q=\allnomagmaq$ is also stronger than the effective $q=0.6$ underlying ZC17's estimate of planets' cumulative XUV history (see \S\ref{s:physics}), as seen by the wider spacing between lines of different $L_\star$. Relative to ZC17, the new inferred shoreline is more generous to hot planets around hot stars, but similarly harsh to planets with Earth-like bolometric fluxes around the coolest stars.

\begin{figure}
    \centering
    \includegraphics[width=\columnwidth]{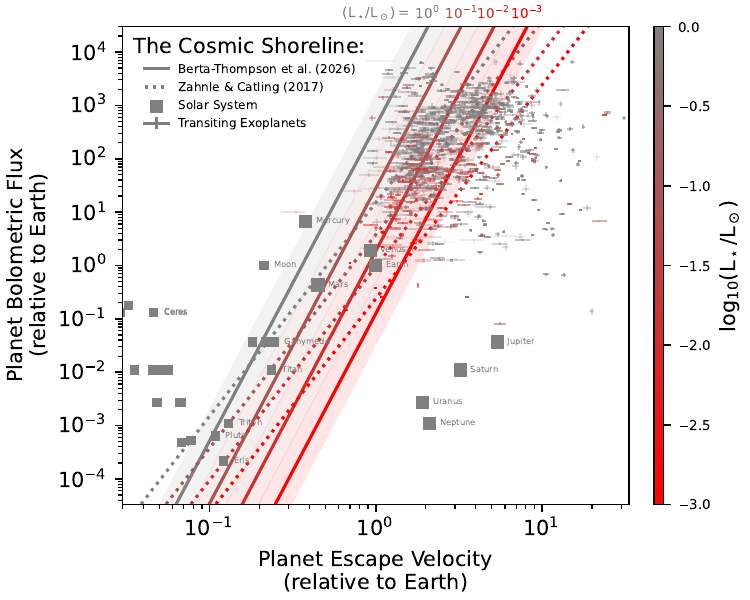}
    \includegraphics[width=\columnwidth]{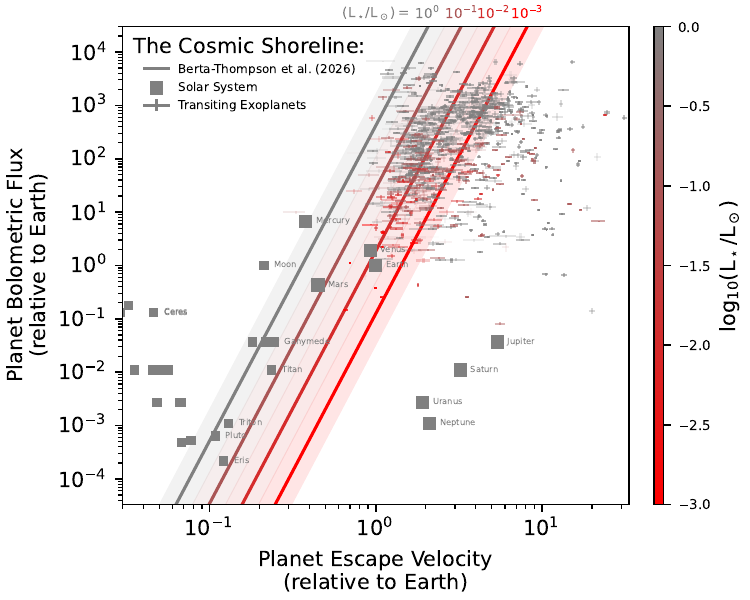}
    \caption{The inferred cosmic shoreline with all planets compressed onto a single plot, both with and without comparison to the traditional \citep{zahnleCosmicShorelineEvidence2017a} XUV-driven shoreline. Colors indicate host star luminosity $L_\star$ both for individual planets and for the shoreline to which they should be compared, with swaths spanning the $\pm 1w$ intrinsic logistic width of the shoreline. Plotted lines follow the equation $\log_{10} (f_{\rm shoreline}/f_\earth) = \allowbreak \log_{10} (f_{\sf 0}/f_\earth) +\allowbreak p \log_{10} (v_{\sf esc}/v_{\sf esc, \earth}) +\allowbreak q \log_{10} (L_\star/L_\sun)$. Although we encourage future users to incorporate uncertainties by drawing samples from the parameter posteriors, the best-fit parameters used here were $\log_{10} (f_{\sf 0}/f_\earth) = \allnomagmalogfojustvalue$, $p = \allnomagmapjustvalue$, $q=\allnomagmaqjustvalue$, and $\ln w = \allnomagmalnwjustvalue$ (\href{https://github.com/zkbt/shoreline/blob/main/notebooks/8-plot-how-shorelines-compare-to-zc17.ipynb}{\texttt{</>}}).
    }
    \label{f:shoreline-zc17}
\end{figure}

\subsection{Predictions for JWST Rocky Worlds}
\label{s:implications-rocky-worlds}

What does this new shoreline model predict for the atmospheres that the JWST Rocky Worlds DDT program may find?
Table \ref{t:rocky-worlds} and Figure \ref{f:shoreline-rocky} show estimated atmosphere probabilities for the 9 planned Rocky Worlds targets. Of these, GJ 3929 b has already been observed, and its atmospheric non-detection \citep{2025_xue_JWSTRockyWorlds} was included in the shoreline fit. LTT 1445Ab's bright MIRI/LRS spectrum \citep[][where we confirmed the near-zero eccentricity of the orbit]{wachiraphanThermalEmissionSpectrum2025a} also contributed to the current shoreline, but the Rocky Worlds' 15$\mu$m photometry can probe much more tenuous CO$_2$ atmospheres \citep{ihConstrainingThicknessTRAPPIST12023} and could in the future flip it to $A_{\sf i}=1$. Figure \ref{f:shoreline-rocky} shows the Rocky Worlds targets span a broad range of fluxes, gravities, {\em and} host star luminosities, with motivation for such breadth emerging from predictive target optimization experiments \citep{2025_ih_RockyPlanetsStars}. Here, we predict 5/9 of these Rocky Worlds targets to have atmosphere probabilities $>50\%$, and the expected number of atmospheres (= the sum of these probabilities) to be 5.36. By spanning a range of shoreline distances ($\Delta_{\sf i}$, see Equation \ref{e:delta}), the new Rocky Worlds observations will have strong leverage to test the predictions of this shoreline model and update its parameters, in addition to discovering what will likely be some of the easiest-to-observe rocky exoplanet atmospheres in the Universe.

\begin{deluxetable*}{lrrrr}[ht]
\tablecaption{Predicted atmosphere probabilities for JWST Rocky Worlds Targets}

\tablehead{\colhead{Planet} & \colhead{$\log_{10}(f/f_\oplus)$} & \colhead{$\log_{10}(v_{\rm esc}/v_{esc,\oplus})$} & \colhead{$\log_{10}(L_\star/L_\odot)$} & \colhead{$P(A_{\sf i} =1)$}}
\startdata
GJ-3929 b & 1.287 & 0.023 & -1.965 & $4.53 \pm 3.8\%$ \\
LTT-1445 A c & 1.066 & 0.048 & -2.056 & $10.4_{-7.6}^{+8.3}\%$ \\
LTT-1445 A b & 0.756 & 0.158 & -2.091 & 7$7.1_{-14}^{+13}\%$ \\
LHS-1140 b & -0.424 & 0.255 & -2.470 & $99.3_{-0.27}^{+0.67}\%$ \\
TOI-198 b & 0.485 & 0.217 & -1.509 & $99.7_{-0.26}^{+0.29}\%$ \\
TOI-406 c & 1.251 & 0.099 & -1.712 & $43.1_{-16}^{+15}\%$ \\
TOI-771 b & 1.248 & 0.160 & -2.260 & $21.6_{-14}^{+14}\%$ \\
HD 260655 c & 1.202 & 0.153 & -1.441 & $91.8 \pm 4.6\%$ \\
TOI-244 b & 0.865 & 0.123 & -1.651 & $88.6_{-6.7}^{+6.5}\%$
\enddata
\label{t:rocky-worlds}
\tablecomments{Planet inputs are calculated from the table on the JWST Rocky Worlds project website (\url{https://rockyworlds.stsci.edu}), retrieved 1 March 2026. Atmosphere probabilities $P(A_{\sf i}=1)$ are calculated via Equation \ref{e:p_i}, sampling from the parameter posterior distribution (but not the individual planet property uncertainties) and quoting central 68.3\% confidence intervals on the predicted probability (\href{https://github.com/zkbt/shoreline/blob/main/notebooks/10-apply-to-rocky-worlds.ipynb}{\texttt{</>}}).}
\end{deluxetable*}

\begin{figure*}[ht]
    \centering
    \includegraphics[width=\textwidth]{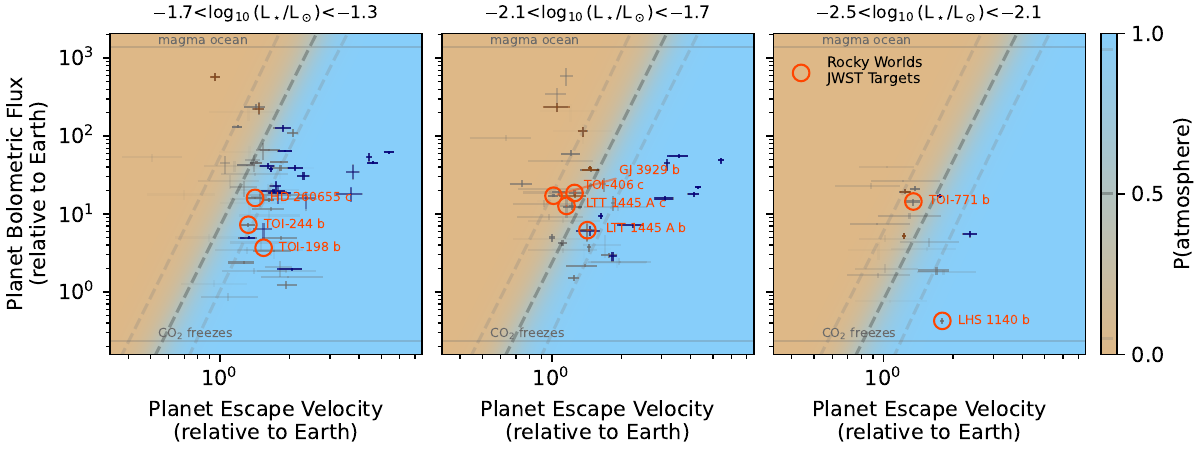}
    \caption{Cosmic shoreline predictions highlighting the JWST Rocky Worlds targets being observed in thermal emission at 15 $\mu$m with JWST/MIRI. These panels are strongly zoomed relative to Figures \ref{f:shoreline}, \ref{f:shoreline-animated}, and \ref{f:shoreline-CO2} to show detail on the positions of each planet relative to the shoreline. By spanning a broad range of stellar luminosity, escape velocity, and bolometric flux, the Rocky Worlds planets will provide sensitive tests of the shoreline model proposed here and/or refinement to its parameters (\href{https://github.com/zkbt/shoreline/blob/main/notebooks/10-apply-to-rocky-worlds.ipynb}{\texttt{</>}}).}
    \label{f:shoreline-rocky}
\end{figure*}

\subsection{Relevance for Habitable Zone Planets}
\label{s:implication-hz}
Can all planets in the habitable zone retain atmospheres? 
Cast in terms of semimajor axis $a = \sqrt{L_\star/4\pi f}$, the shoreline orbital distance scales as $a_{\sf shoreline} \propto L_\star^{(1-q)/2}$. Our inferred $q = \allnomagmaq$ is mostly $> 1$, implying that $a_{\sf shoreline}$ grows {\em outward} toward smaller stars. In contrast, the habitable zone distance necessarily shrinks inward toward smaller stars as $a_{\sf hz} \propto L_\star^{1/2}$ \citep[generally making them easier to observe;][]{blakeNearInfraredMonitoringUltracool2008, nutzmanDesignConsiderationsGroundBased2008a}. Figure \ref{f:shoreline-hz} demonstrates where these two curves intersect. For these plots, we use the mass-radius relation of Figure \ref{f:mass-radius} to translate $v_{\sf esc}$ into planet radius $R_{\sf p}$, and we show predictions for roughly Earth-size planets as well as for larger super-Earths.

The very idea of the habitable zone depends on a planet retaining an atmosphere to support a hydrological cycle and climate feedbacks. If a planet falls within the classical habitable zone but also on the dry side of the cosmic shoreline, the planet has likely lost its atmosphere, its water, and its habitability. With the shoreline model presented here (and also the original from ZC17), habitable-zone planets orbiting the lowest mass M dwarfs are likely unable to retain atmospheres. Though such planets offer enormous observational advantages for transit measurements, the harsh XUV and atmospheric loss environment of low-mass stars suggests we may have more luck finding atmospheres on slightly larger planets orbiting slightly hotter stars.

\begin{figure*}
    \centering
    \includegraphics[width=\textwidth]{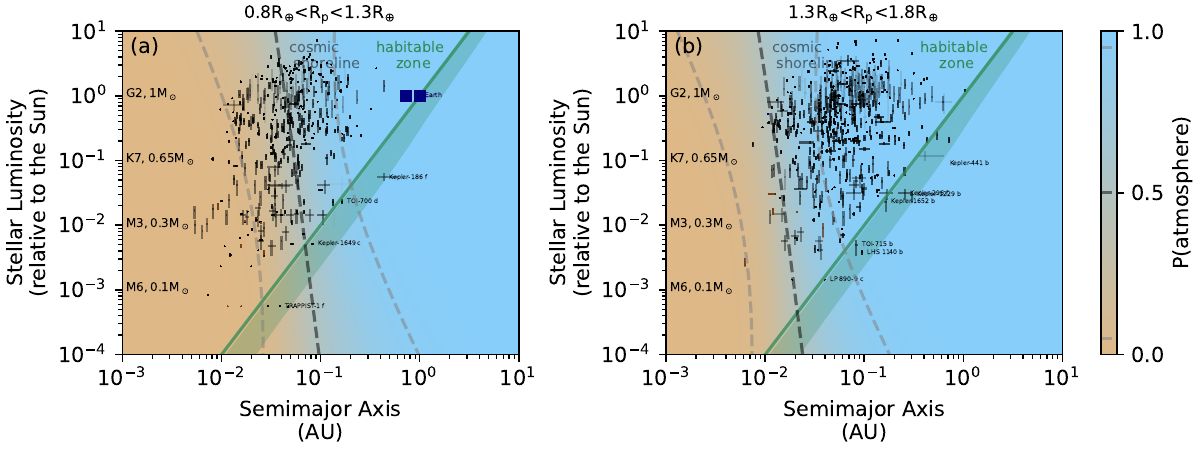}
    \caption{The cosmic shoreline model from this paper, transformed into planet semimajor axis $a$ and planet radius $R_{\sf p}$, and and compared to liquid-water habitable zone. Approximate main-sequence spectral types and stellar masses are provided for reference. Smaller Earth-size rocky planets (a) are less able to retain atmospheres than larger super-Earths (b), such that Earth-size planets in the habitable zone of the lowest mass stars have likely lost their atmospheres (\href{https://github.com/zkbt/shoreline/blob/main/notebooks/7-plot-shoreline-compared-to-hz.ipynb}{\texttt{</>}}).}
    \label{f:shoreline-hz}
\end{figure*}
}

\section{Conclusions}
\label{s:conclusions}

In this paper, we present a probabilistic 3D cosmic shoreline model that defines the maximum bolometric flux $f_{\sf shoreline}$ a planet with given escape velocity $v_{\sf esc}$ and stellar luminosity $L_\star$ can receive and still maintain a substantial atmosphere. We infer parameters for this model by fitting to exoplanets and Solar System bodies with atmospheres or global surface volatiles. We currently see no strong evidence for different shoreline parameters when we zoom in to consider only temperate atmospheres where CO$_2$ can exist as a gas, just larger uncertainties on the shoreline parameters. 

We provide three tools to help reproduce, expand, and/or make use of the results presented in this paper: 
\begin{itemize}
    \item The \texttt{exoatlas} Python code to access, filter, and visualize archival planet properties \citep{berta-thompsonZkbtExoatlas2025}, which is publicly available via \texttt{pip install exoatlas} and under review at the Journal of Open Source Software.
    \item \texttt{jupyter} notebooks to reproduce all paper figures and major analyses in a GitHub repository (\href{https://github.com/zkbt/shoreline}{\texttt{</>}}). 
    \item a \texttt{zenodo} repository containing  parameter posterior samples for the main cosmic shoreline fit, as well organized planet populations, additional test posteriors, and a more expansive collection of figures and animations \citep{2026_berta-thompson_3DCosmicShoreline}.
\end{itemize}

The 3D shoreline presented here (Equations \ref{e:log_f_shoreline} + \ref{e:p_i}) relies on four parameters: $\log_{10} (f_0/f_\oplus) = \allnomagmalogfo$, $p=\allnomagmap$, $q = \allnomagmaq$, and $\ln w = \allnomagmalnw$. Using the multivariate posterior probability distribution of these parameters, we can predict answers to a few basic hypothetical questions about our own Solar System (\href{https://github.com/zkbt/shoreline/blob/main/notebooks/9-apply-shoreline-to-specifc-questions.ipynb}{\texttt{</>}}):
\begin{itemize}
\item How much bigger would Mercury need to be to retain an atmosphere? Orbiting the Sun at 0.39 AU and receiving $f/f_\Earth = 6.7$, Mercury would need an escape velocity of at least $v_{\sf esc}/v_{\sf esc, \earth} = (f/f_{\sf 0})^{1/p} = \applymercuryescapevelocity$ to have a 50\% chance of having an atmosphere. By the mass-radius relation in Figure \ref{f:mass-radius}, this translates to about $\applymercuryr~ \mathrm{R_\earth}$, or $\applymercuryrratio \times$ its current size. 

\item How much hotter could Venus be before losing its atmosphere? Moving this approximately Earth-size planet with $v_{\sf esc}/v_{\sf esc, \earth} = 0.93$ inward to the Sun would apparently permit it to retain significant atmosphere until it reaches the shoreline at $\log_{10} (f/f_{\earth}) = \applyvenuslogflux$, or $a = \applyvenussemimajor$ AU. This suggests ultra-close rocky planets around FGK stars may be able to retain atmospheres, as difficult as they may be to observe.

\item How much could we shrink the Sun's mass/radius/luminosity before a habitable-zone Earth-size planet can no longer maintain any atmosphere at all? For 1$\mathrm{R_\earth}$ planets with $v_{\sf esc}/v_{\sf esc, \earth} = 1$, the cosmic shoreline intersects with $f/f_\Earth =1 $ at $\log_{10}(L_\star/L_\sun) = \applyearthlogluminositylimit$ (or roughly M4 spectral type or 0.25 $\mathrm{M_\sun}$ mass). Notably, for larger $1.5 \mathrm{R_\earth}$ planets with $v_{\sf esc}/v_{\sf esc, \earth} \approx 1.6$, this intersection extends down to $\log_{10}(L_\star/L_\sun) = \applyearthandahalflogluminositylimit$ (roughly M8V spectral type or 0.09 $\mathrm{M_\sun}$ mass, approximately TRAPPIST-1).

\item How much must we perturb planets to move them from one side of the shoreline to the other? We include an intrinsic width to the shoreline, finding that transitioning from a 95\% chance of having an atmosphere to a 95\% chance of not having one spans  $w_{95} = \applywninefivef$ dex in bolometric flux, $\applywninefivev$ dex in escape velocity, or $\applywninefiveL$ dex in stellar luminosity. This intrinsic width exceeds the measurement uncertainties for most well-characterized transiting planets, so they have little effect on the inferred shoreline parameters.

\end{itemize}

Of the 9 rocky planets being observed in the JWST Rocky Worlds DDT program, we predict 5 have a $>50\%$ chance of hosting a detectable atmosphere. We caution that the atmosphere labels in this paper are a still little fuzzy, with ``no atmosphere'' often really meaning ``probably $\lesssim 10$ bar CO$_2$''. Rocky World's 15 $\mu$m MIRI photometry can be sensitive to more tenuous CO$_2$ atmospheres than were detectable for many of the planets used here, potentially converting planets currently labeled as atmosphereless into ones with atmospheres. Other rocky exoplanet JWST programs, like the large Charting the Cosmic Shoreline (JWST-GO-7073) transmission spectroscopy program, will hopefully also add new atmosphere detections in the years to come. This model can be updated as new atmosphere constraints come in, sharpening the inferred shoreline parameters and hopefully improving its predictive and explanatory power. In the meantime, the steep slopes $p$ and $q$ found here imply that future searches for rocky planet atmospheres might be most fruitful for larger rocky planets (higher $v_{\sf esc}$) orbiting more massive stars (higher $L_\star$).

\added{In this work, we attempted to capture the most important first-order inputs for long-term atmospheric evolution: the energy planets receive from their star, the strength of gravity to resist escape, and the knowledge that lower-mass stars emit more of their energy in the XUV wavelengths that matter most for escape. We assume the boundary in this 3D space between worlds with and without atmospheres is planar, with constant log-linear slopes. These are clearly rough approximations. Additional inputs surely matter, such as age, disk chemistry, dynamical history, impact and accretion history, and interior-atmosphere interactions. 

Thinking of how to improve this model, the $\left(v_{\sf esc}/v_{\sf esc, \earth}\right)^p$ term in Equation \ref{e:f_shoreline} could be replaced with more nuanced probabilistic functions based on atmospheric evolution models that explicitly track time-integrated sources and sinks. The $\left(L_\star/L_\sun\right)^q$ term could be replaced with more precise  empirical/theoretical estimates of how XUV and other drivers of atmospheric loss scale with stellar luminosity, with age, and with metallicity. But for now, the fairly broad intrinsic width $w$ we infer for this cartoonish shoreline likely encapsulates errors in the shape of the shoreline, unmodeled hidden variables not yet included in the model, and even potentially an underlying chaotic stochasticity, where two planets with seemingly very similar environments might have diverging atmospheric histories. Empirically disentangling a wiggly shoreline topography or the need for more dimensions from this true randomness will likely require larger sample of rocky exoplanet atmospheres.}

\begin{acknowledgments}
\added{We gratefully thank Autumn Stephens, Valerie Arriero, Mirielle Caradonna, Jackson Avery, Girish Duvvuri, Sebastian Pineda, Yuta Notsu, Dave Brain, Ren\`e Doyon, Nicolas Cowan, Casey Brinkman, Andrew Cumming, the CU Stars and Planets Research Club, and the very thoughtful anonymous referee for conversations that improved this work.} We also thank Dan Foreman-Mackey and Jake VanderPlas for pedagogical statistics blog posts that inspired some of the methods used here. This research has made use of the NASA Exoplanet Archive, which is operated by the California Institute of Technology, under contract with the National Aeronautics and Space Administration under the Exoplanet Exploration Program, as well as the planetary body archive maintained by the JPL Solar System Dynamics group. This material is based upon work supported by the National Science Foundation under Grant No. 1945633, as well as  program \#JWST-GO-2708 provided by NASA through a grant from the Space Telescope Science Institute, which is operated by the Association of Universities for Research in Astronomy, Inc., under NASA contract NAS 5-03127. 
\end{acknowledgments}

\begin{contribution}
Z. Berta-Thompson planned the project, did the analyses, and wrote the manuscript. P. Wachiraphan and C. Murray contributed expertise and reviewed the manuscript.

\end{contribution}

\facilities{Exoplanet Archive, HST, JWST, Spitzer, Kepler, TESS}

\software{\texttt{astropy} \citep{astropycollaborationAstropyCommunityPython2013, astropycollaborationAstropyProjectBuilding2018a, astropycollaborationAstropyProjectSustaining2022},
\texttt{numpy} \citep{harrisArrayProgrammingNumPy2020a}, \texttt{matplotlib} \citep{hunterMatplotlib2DGraphics2007}, \texttt{jax} \citep{jax2018github}, \texttt{numpyro} \citep{phanComposableEffectsFlexible2019}, \texttt{arviz} \citep{arviz_2019}, \texttt{exoatlas} (\href{https://github.com/zkbt/exoatlas}{github.com/zkbt/exoatlas})}

\bibliography{shoreline}{}
\bibliographystyle{aasjournalv7}

\end{document}